\documentclass[twocolumn,10pt]{aastex631}
\usepackage{graphicx}%Include figure files
\usepackage[normalem]{ulem}
\usepackage{rotating}

\usepackage{natbib}
\usepackage{longtable}
\usepackage{amsmath}
\usepackage{lipsum}
\usepackage{amssymb}
\usepackage{nccmath}
\usepackage{cuted}
\usepackage{mathtools}
\usepackage{subfigure}
\usepackage{multirow}
\usepackage{url}
\usepackage{array}
\usepackage[toc,page]{appendix}
\defcitealias{cosep}{COSEP}
\defcitealias{lsst}{I19}
\defcitealias{frontpaper}{B21}
\newcommand{\new}[1]{{\color{black} #1}}

\newcommand{\maf}{\texttt{MAF}}

\newcommand{\eg}{\emph{e.g.}}
\newcommand{\ie}{\emph{i.e.}}
\newcommand{\opsim}{\texttt{OpSim}}

\newcommand{\frackn}{$77\%$}
\newcommand{\fracgrp}{$53\%$}

\newcommand{\GWhlvk}{$7.4^{+11.3}_{-5.5}$}
\newcommand{\GWhlvkf}{100\%}

\newcommand{\KNy}{$4.9^{+7.6}_{-3.6}$}
\newcommand{\KNz}{$5.6^{+8.6}_{-4.1}$}
\newcommand{\KNi}{$5.6^{+8.6}_{-4.1}$}
\newcommand{\KNr}{$5.7^{+8.7}_{-4.2}$}
\newcommand{\KNg}{$5.7^{+8.7}_{-4.2}$}
\newcommand{\KNu}{$5.7^{+8.7}_{-4.2}$}
\newcommand{\KNyf}{76\%}
\newcommand{\KNzf}{76\%}
\newcommand{\KNif}{\frackn}
\newcommand{\KNrf}{\frackn}
\newcommand{\KNgf}{\frackn}
\newcommand{\KNuf}{\frackn}

\newcommand{\deepKNy}{$5.7^{+8.7}_{-4.2}$}
\newcommand{\deepKNz}{$5.7^{+8.7}_{-4.2}$}
\newcommand{\deepKNi}{$5.7^{+8.7}_{-4.2}$}
\newcommand{\deepKNr}{$5.7^{+8.7}_{-4.2}$}
\newcommand{\deepKNg}{$5.7^{+8.7}_{-4.2}$}
\newcommand{\deepKNu}{$5.7^{+8.7}_{-4.2}$}
\newcommand{\deepKNyf}{\frackn}
\newcommand{\deepKNzf}{\frackn}
\newcommand{\deepKNif}{\frackn}
\newcommand{\deepKNrf}{\frackn}
\newcommand{\deepKNgf}{\frackn}
\newcommand{\deepKNuf}{\frackn}

\newcommand{\GRBoptical}{$0.15^{+0.22}_{-0.11}$}
\newcommand{\GRBopticalf}{2\%}
\newcommand{\deepGRBoptical}{$0.37^{+0.56}_{-0.27}$}
\newcommand{\deepGRBopticalf}{5\%}

\begin{document}

\title{Kilonova parameters estimation with LSST at Vera C. Rubin Observatory}

\author[0000-0003-2132-3610]{Fabio Ragosta}
\affil{INAF, Osservatorio Astronomico di Roma, via Frascati 33, I-00078
Monte Porzio Catone (RM), Italy}\altaffiliation{LSSTC Data Science Fellow}
\affil{Space Science Data Center – ASI, Via del Politecnico SNC, 00133 Roma, Italy}
\author[0000-0002-2184-6430]{Tom{\'a}s Ahumada}\altaffiliation{LSSTC Data Science Fellow}
\affil{Department of Astronomy, University of Maryland, College Park, MD 20742, USA}
\affiliation{Astrophysics Science Division, NASA Goddard Space Flight Center, MC 661, Greenbelt, MD 20771, USA}
\affil{Center for Research and Exploration in Space Science and Technology, NASA Goddard Space Flight Center, Greenbelt, MD 20771, USA}

\author[0000-0002-8875-5453]{Silvia Piranomonte}
\affiliation{INAF, Osservatorio Astronomico di Roma, via Frascati 33, I-00078
Monte Porzio Catone (RM), Italy}

\author[0000-0002-8977-1498]{Igor Andreoni}
\altaffiliation{Neil Gehrels Fellow}
\affiliation{Joint Space-Science Institute, University of Maryland, College Park, MD 20742, USA}
\affil{Department of Astronomy, University of Maryland, College Park, MD 20742, USA}
\affiliation{Astrophysics Science Division, NASA Goddard Space Flight Center, MC 661, Greenbelt, MD 20771, USA}
\author[0000-0002-2810-2143]{Andrea Melandri}
\affiliation{INAF, Osservatorio Astronomico di Roma, via Frascati 33, I-00078
Monte Porzio Catone (RM), Italy}

\author[0000-0002-7439-4773]{Alberto Colombo}
\affil{Università degli Studi di Milano-Bicocca, Dipartimento di Fisica “G. Occhialini”, Piazza della Scienza 3, I-20126 Milano (MI), Italy} 
\affiliation{INFN–Sezione di Milano-Bicocca, Piazza della Scienza 3, I-20126 Milano (MI), Italy}
\affiliation{INAF–Osservatorio Astronomico di Brera, via E. Bianchi 46, I-23807 Merate (LC), Italy}

\author[0000-0002-8262-2924]{Michael W. Coughlin}
\affiliation{School of Physics and Astronomy, University of Minnesota, Minneapolis, Minnesota 55455, USA}

\begin{abstract}

The up-coming Vera Rubin Observatory's Legacy Survey of Space and Time (LSST) opens a new opportunity to rapidly survey the southern Sky at optical wavelenghts (\ie ugrizy bands).
In this study, we aim to test the possibility of using LSST observations to constrain the mass and velocity of different KN ejecta components from the observation of a combined set of light curves from GRB afterglows and KNe. We used a sample of simulated light curves from the aforementioned events as they would have been seen during the LSST survey
to study how observing strategies' choice impacts the parameters' estimation. 
We found that the observing strategy design to be the best compromise between light curves coverage, observed filters and fit reliability involves a high number of visits with long gap pairs of about 4 hours every 2 nights in the same or different filters. The features of the observing strategy will allow us to recognize the different stages of the light curve evolution and gather observations in at least 3 filters.
\end{abstract}
\section{Introduction}

The detection of GW170817, a binary neutron star (BNS) merger, using both gravitational waves (GW) and photons, marked a groundbreaking milestone in multi-messenger astronomy. Initially, GW170817 was identified solely by its gravitational wave signal \citet{2017AbbottGW, Abbott_2017c}, and subsequently, an array of electromagnetic (EM) signals from ground-based and space-borne telescopes covering the entire EM spectrum confirmed the presence of a luminous electromagnetic counterpart to the event  \citet{2017AbbottMMA}.
In particular, approximately 11 hours after the GW detection, the search for the EM signal of GW170817 led to the discovery of an electromagnetic counterpart named AT2017gfo associated with the GW signal \citet{2017Pian, 2017Coulter,2017Soares, Arcavi:2017xiz, Andreoni:2017ppd, Lipunov_2017,2017Valenti, 2017Smartt, Abbott_2017c, Buckley2017,Cannon_2012, Diaz_2017, Drout2017,Evans2017,HU2017,Kilpatrick2017,McCully_2017,2017Troja,2017Utsumi,2017Tanvir,2017Valenti}.

This discovery played a crucial role in addressing numerous issues in high-energy astrophysics and fundamental physics. It significantly contributed to resolving the origins of short $\gamma$-ray bursts (sGRBs), the existence of kilonovae ($KNe$), and the processes behind heavy element synthesis. 
Additionally, it offered valuable independent constraints on two key aspects of astrophysics. First, it provided insights into the previously unknown equation of state (EoS) of neutron stars (NS), as discussed in \citet{2018Abbott}. It also helped refine the understanding of the Hubble constant \citet{2017AbbottHubble, 2018Cantiello, kashyap2019can, 2019Fishbach, 2019Hotokezaka, 2020Coughlin, doctor2020thunder, 2020Dietrich, coughlin2020standardizing, 2021AbbottH0}.
These findings represent significant steps forward in our comprehension of the cosmos and have opened new avenues for future research in these fields.

The concept of KNe as transient phenomenon powered by the radioactive decay of synthesized heavy $r$-process elements resulting from the ejection of neutron star matter during compact mergers, was first highlighted by \citet{1974Lattimer}. Several subsequent studies have contributed to our understanding of this event, including \citet{1998Pac}, \citet{Freiburghaus_1999}, \citet{2000Lattimer}, \citet{2010Metzger}, \citet{2011Roberts}, \citet{2013Tanaka}, \citet{2014Metzger}, \citet{2014Grossman}, \citet{2015Kasen}, and \citet{2016Barnes}.
Along with the ejection of neutron-rich material, a relativistic jet is also produced, the jet, moving close to the speed of light, emits a powerful beam of $\gamma$-ray radiation, leading to a so called sGRB prompt emission. 
As the jet interacts with the interstellar medium, it decelerates and produces a detectable afterglow powered by synchrotron emission, observable from X-rays to radio frequencies.
Before the detection of GRB 170817A, the connection between sGRBs and compact object mergers had been supported only by indirect evidence \citet{2017Goldstein, 2017Savchenko,Abbott_2017c,Tanvir2013,Fong2015}. However, the simultaneous detection of GW and $\gamma$-rays demonstrated that at least a portion of sGRBs are indeed associated with the merging of binary neutron stars.
Currently, the conventional scheme for the progenitors of GRBs is a subject of debate, as counterexamples have emerged in recent years \cite{ahumada2021,rastinejad2022, mei2022gigaelectronvolt, yang2022long, troja2022nearby}. To complicate matters further, indirect detections of KN emission have been proposed, supported by the identification of optical and near-infrared (NIR) excesses in the flux of some GRB afterglows \citep{Tanvir2013,troja2019afterglow, 2020NJin, 2020Rossi,rastinejad2022}.

Constraining the properties of these particular events is crucial in resolving the ongoing debate regarding the dominant site for the production of r-process nuclei in the universe. Some studies \citep[e.g.,][]{2017Kasen,anand2023} argue that binary neutron star mergers are the primary source, while others \citep{2019SiegelGW} suggest the collapse of massive stars as the main contributor.

sGRB detection rates range between 10 and 40 per year, for the GRB instruments on board of the Neil Gehrels \textit{Swift} Observatory \citep{swift}
and the \textit{Fermi} satellite respectively \citep{fermi2020cat}.
However, the optical counterparts for these bursts have proven to be elusive, mainly because the localization of \textit{Fermi} sGRBs typically spans hundreds of square degrees e.g. \citep[][]{ahumada2022, mong2021searching}. The follow up of BNS and neutron star--black hole (NSBH) mergers detected by the International Gravitational-Wave Network  (IGWN), consisting of Advanced LIGO, Advanced Virgo, and KAGRA during the third observing run (O3) has not been fruitful, possibly due to the fact that the GW skymaps are similarly large \citep{Goldstein19gw,andreoni2019growth,coughlin2019growth,Andreoni20gw,kasliwal2020kilonova,2020Gompertz,2021Chang,2022Petrov}.
The lack of counterpart detections can therefore be explained by the fast fading nature of 
both KNe and afterglows, 
the large skymaps to observe, and the low local rate of compact binary mergers \citep{2020Dichiara}.

Empirical constrains on KN rates by optical surveys set an upper limit to $R<900~\frac{Gpc^3}{yr}$ \citep{Andreoni_2020, 2021Andreoni} for KNe\ similar to AT2017gfo. 
Moreover, only a fraction of those will be detectable, as they could be beyond the detection limit of available telescopes. Obscuration and absorption by the Galactic plane is also a significant limitation to the detection of counterparts. 
In light of these constraints, the expected BNS detection rate of $4-80$ events per year for the LVK network after 2020 \citep{2018Abbott_obs, 2022Petrov}, based only on GW searches, will likely provide only a few tens of detections throughout the next decade.

The new Legacy Survey of Space and Time \citep[LSST, ][]{2009LSST} is expected to be a game-changing facility in astrophysics. Time-domain astronomy will particularly benefit from the large $\sim 10~ \mathrm{deg}^2$ field of view (FoV) of the camera combined with the depth achievable with the 8.4 m-diameter primary mirror, with an effective aperture of 6.423 m. Depending on the choice of the LSST cadence, the project can unveil a large number of KNe\ and other types of fast fading transients \citep [\eg ][]{2022AndreoniA}.

The current estimates of KNe rates tell us that LSST will be able to detect $\approx 10^2 - 10^3$ events within $z = 0.25$ during the entire survey \citep{2018Valle}. 
Moreover, \citet{2022AndreoniA} demonstrated that LSST is expected to find more than 300 KNe\ out to $\approx 1400$ Mpc over a ten-year survey. 
Among those, we expect about $3-32$ KNe recognizable as fast-evolving transients similar to the one associated to GW170817. Furthermore, KNe have been analyzed only in association with other events like GRB or GW detection, thus the possibility to detect and recognize such events is strictly related to the ability to survey as fast as possible the wide error boxes from GW signals and, once located, to promptly analyze their EM emission.

In spite of the technological and instrumental advances across multiple wavelengths, the fast evolving nature of KNe will likely impede their spectral analysis. For this reason, this paper aims to analyze multiple observational strategies that only rely on photometry to derive physical parameters  of a KN sources without using spectra. Throughout this paper we assume that we know the location and the energy of the merger from other messengers (GW and GRB). 

\new{We consider that KNe can potentially be detected as a possible additional component to the optical and NIR afterglow of short GRBs in the temporal windows that goes from few hours to a few weeks after the onset of the burst \citet[\eg][]{2015Kasen, 2016Barnes, 2016Fern, 2017Metzger}.
This assumption follows the findings of \citet{2020Rossi} where the authors were able to isolate a golden sample of GRB afterglows which behaviour indicates the presence of a KN component in the afterglow light curves. However, they stated that high constraints on redshift or NIR observation are needed to be able to find KN contribution in the afterglow. 
In this paper we used a similar method to \citet{2020Rossi} for comparing GRB afterglows lightcurves with KNe in all the LSST observable bands to define an optimal observing strategy that can enhance the ability of detect KNe\ and characterize their sources, even using their indirect observations through GRB afterglows.
}

The paper is organized as follows: in \autoref{sec:the_need_for} is described the impact of parameter estimation to understand the physics behind the KN explosion; in \autoref{sec:method} the methodology of the simulations and analysis of the extracted data is described; in \autoref{sec:MAF} the observing strategies are summarized; \autoref{sec:performance} and \autoref{sec:model_impact} are dedicated to the analysis of the model and observing strategies' features which impact the parameters estimation; in \autoref{sec:param_estim} and \autoref{sec:opsim} are dedicated to the description of the obtained results; finally \autoref{sec:conclusion} is dedicated to the discussion of the results and the conclusion. 

\section{The need for parameters estimation}\label{sec:the_need_for}
\label{sec:the_need_for}

\begin{figure}
    \centering   
    \hspace*{-1cm}           
    \includegraphics[scale=.2]{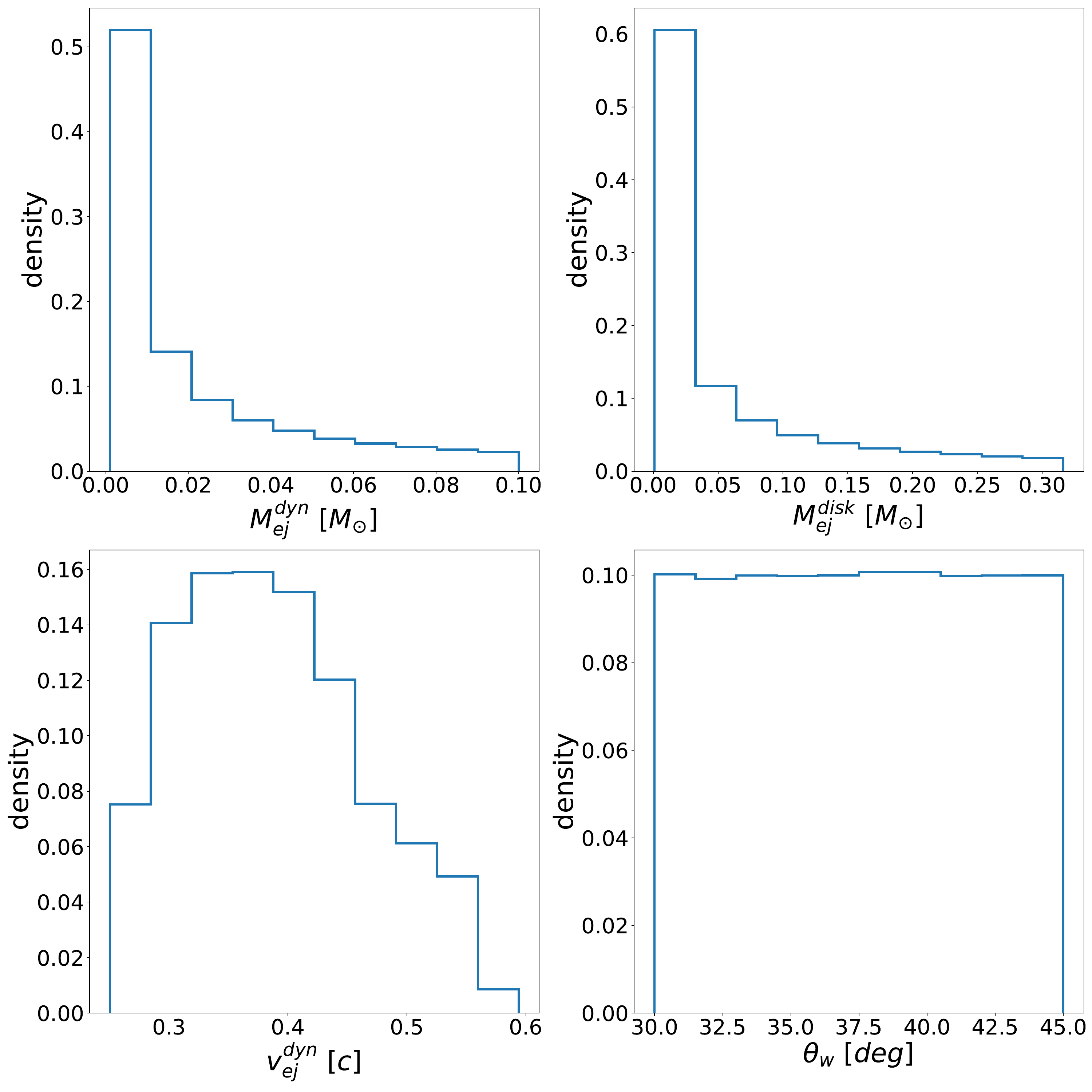}
    \caption{The distribution of the model parameters we injected in \texttt{nmma} to simulate KN light curves, the ejecta mass and velocity distribution is modeled by the binary mass distribution in \citep{Dietrich20},
    the open angle for the wind ejecta component is drawn by a uniform distribution 
    $\mathcal{U}(\pi/6,\pi/4)$.}
    \label{fig:apriorimass}
\end{figure}

\new{KNe observations encode information on both ejecta properties and ejection mass processes that happen during the merger and post-merger \citep{2017Metzger,2018Coughlin,2017Villar}. Part of the NS (whether the system is a BNS or a NSBH) can be expelled and become unbound \citep{Davies:1993zn,Rosswog:1998hy}. Tidal forces right before the merger can cause partial disruption of the NSs (or of the single NS in the case of the NSBH) with material lunched at mildly relativistic velocities on the orbital plane of the system. Once the accretion disk is formed, neutrino radiation and nuclear recombination with magnetodynamic viscosity can drive mass outflow from the disk \citep[see ][ and reference therein]{2020Ascenzi}. The velocity, the mass and the geometry of the ejecta strictly depend on the properties of the system involved, thus, a statistical study of a population of KNe would help us in understanding the distribution of those parameters.}

Several studies of GW170817 attempted to infer the amount of material ejected \citep[e.g.,][]{2017Metzger, 2018Coughlin,2021Breschi,2021Heinzel,2019Coughlin,2017Smartt,2017Tanvir,2017Alexander,2017Chornock,2017Cowperthwaite,2022Collins,2022Ristic}. 
However, the extent and the properties of the ejected material from this event remain uncertain and in considerable tension with theoretical expectations for the amount of each type of ejecta component\citep{2022Collins,2021Nedora,2018Radice,2021Korobkin}, likely in part because of underestimated uncertainties in these theoretical ejecta evaluations \citep{2022Henkel}.
\new{KNe are rare and faint events compared to supernovae and other common classes of extragalactic transients, so they are hard to detect. Observational constraints, due to their intrinsic peculiarity, impact on our understanding of the processes the source undergoes. This implies we are not able to grasp the complexity of the compact object merger without making some important simplification, such as assuming a simplified treatment of the radiative transport that overlooks detailed three-dimensional anisotropic considerations. Similarly, opacity is considered in a simplified manner, without taking into account sophisticated nuclear reaction networks and composition variations \citep[\eg][]{2018Wollaeger,2021Wollaeger}.}
Recent calculations using improved anisotropic radiative transfer and opacity calculations still arrive at similar conclusions for the AT2017gfo event description \citep{2017Metzger,2021Heinzel,2022Ristic,2022Collins,2021Almualla}: a relatively high mass of the blue component  expels from the poles while significant mass of the red component is preferentially ejected toward the equator. 

In particular, 
\citet{Ascenzi19} show the posterior distribution of the KN parameters (ejecta mass, velocity, and lanthanides content) extracted from the observed multi-wavelength afterglow light curves, assuming that the GRB afterglow light curve excess flux can be related to the radioactive decay from the KN ejecta. \new{To estimate the parameters the authors used a KN and GRB afterglow joined model to reproduce the observed data, they produced a statistical distribution of the value from a sample of observed GRB afterglow light curves they claimed to be associate to KN event. The uncertainties on the derived parameters are considerable, primarily because of the sparsely populated light curves and the incomplete understanding of the nuclear reaction network responsible for producing the KN component. As a consequence of these uncertainties, certain cases result in parameter distributions that appear nearly uniform, lacking distinct patterns or trends.}

Well constrained parameters will allow us to: 
(i) add features to the theoretical description of the the event or (ii) break degeneracy between models \citep[e.g. data from AT2017gfo agree with both two and three component ejecta model as shown in ][]{Cowperthwaite_2017}. 
\new{The lack of both photometric and spectroscopic data poses a limitation, compelling us to address the challenge of constraining models solely based on photometric data. This situation highlights the opportunity to enhance observing strategies to optimize the chances of verifying theoretical predictions concerning KN events. By improving observational techniques and data collection, we can better assess and validate the theoretical models of these events.}

\subsection{Detection rates in O4}\label{sec:det_rate}
\begin{table*}[htbp]
\centering
\begin{tabular}{c c cccccc c c}
\hline
\multicolumn{1}{c}{\multirow{2}{*}{\textbf{}}} & \multicolumn{1}{c}{GW} & \multicolumn{6}{c}{KN + GW O4} & & \multicolumn{1}{c}{GRB Afterglow+KN+GW O4} \\\cline{3-8}
& HLVK O4 & \textit{y} & \textit{z} & \textit{i} & \textit{r} & \textit{g} & \textit{u} & & Optical \\ \hline
Shallow Limit & 12 & 22.1 & 23.3 & 24 & 24.7 & 25.0 & 23.9 & & 25 \\ 

Rate & \GWhlvk & \KNy & \KNz & \KNi & \KNr & \KNg & \KNu & & \GRBoptical \\ 
(\% of O4 GW) & (\GWhlvkf) & (\KNyf) & (\KNzf) & (\KNif) & (\KNrf) & (\KNgf) & (\KNuf) & & (\GRBopticalf) \\
\hline

Deep Limit & / & 24.9 & 26.1 & 26.8 & 27.5 & 27.4 & 26.1 & & 27.4 \\ 
Rate & / & \deepKNy & \deepKNz & \deepKNi & \deepKNr & \deepKNg & \deepKNu & & \deepGRBoptical \\   
(\% of O4 GW) & / & (\deepKNyf) & (\deepKNzf) & (\deepKNif) & (\deepKNrf) & (\deepKNgf) & (\deepKNuf) & & (\deepGRBopticalf) \\

\end{tabular}
\caption{Estimation of the expected observation rate of KNe in association with a GW and GRB afterglow (details on how the single values in the table have been estimated in \autoref{TableRate}). Below each rate, we report in parentheses the fraction over the total O4 BNS GW rate (HLVK O4). The GW detection limits refer to the S/N net threshold.
Limiting magnitudes of LSST filters are in the AB system \citep{2009LSST}; detection rates are in units per year.
The reported errors, given at the 90\%  credible level, stem from the uncertainty of the overall merger rate, while systematic errors are not included. 
These results and the underlying methodology are described in 
\autoref{TableRate}.
}

\label{tab:det_rates}
\end{table*}

Rubin Observatory's primary survey, named Wide Fast Deep (WFD), covers an area of $18000$ deg$^2$ through its “universal cadence”, while approximately 10\% of the observing time is reserved for other programs, including intensive observation of Deep Drilling Fields (DDF). Compared to typical points on the sky, the DDF will receive deeper coverage and more frequent temporal sampling in at least some of the LSST Camera's ugrizy filters. 

To estimate the detection rate of EM counterparts potentially detectable by LSST, we constructed a population of KNe\ and GRB afterglows lightcurves, starting from a realistic population of BNSs, following the method described in \cite{2022Colombo, 2023Colombo}, summarized in \autoref{TableRate}. We considered two sets of limiting magnitudes: one shallow related to the expected visits depth of LSST for WFD and one deep related to DDF.  

Assuming a LIGO, Virgo and KAGRA detectors network with the projected O4 sensitivities and a $70\%$ uncorrelated duty cycle for each detector, we find a GW detection rate of \GWhlvk yr$^{-1}$, \frackn\ of which can produce a KN and \fracgrp\ a relativistic jet. The shallow limiting magnitudes for WFD are sufficient to the detect the majority of KNe\ in the $y$ and $z$ bands and all the KNe\ in the $i$, $r$, $g$ and $u$ bands, with a corresponding detection rate of \deepKNg\ yr$^{-1}$. Instead, the fraction of events with a detectable otpical GRB afterglow is between $2-5 \%$, with a maximum detection rate of \deepGRBoptical\ yr$^{-1}$. This is due to the large abundance of off-axis jets within the estimated GW horizon, with a corresponding faintness of these counterparts. In Table \ref{tab:det_rates} we report all the detection rates and the assumed limiting magnitudes. 

\section{Method}\label{sec:method}
\new{Fostered by the discovery of AT2017gfo and its luminous blue emission, several attempts were made to find similar cases in archival short GRB observations \citep[\eg][]{troja2019afterglow, 2020Rossi}. For instance, \citet{troja2019afterglow} found  that some nearby event have optical luminosities comparable to AT2017gfo. In particular, they showed that the sGRB 150101B was a likely analogue to GW170817, characterized by a late peaking afterglow and a luminous optical KN emission, dominating at early times. This finding suggests that KNe\ similar to AT2017gfo could have been detectable in the optical spectrum even though they might not have been explicitly identified before the discovery of GW170817. 
Driven by this motif, 
we aim to study the performance of the KNe parameter estimation to better understand the physical properties of the KN ejecta and their evolution using LSST observational stategies. }
The work is divided by three major parts: 
\begin{enumerate}
    \item simulation of a sample of KN+GRB light curves (\autoref{sec:simulations});
    \item simulation of the observed light curves using a realistic cadence strategy (\autoref{sec:MAF});
    \item estimation of the parameters variance using a bayesian fitting algorithm to retrieve the posterior distribution (see \autoref{sec:performance})
\end{enumerate} 

The parameter estimation of a transient lightcurve, is something that is usually done after the follow up, so the ansatzs here are that we already took care about distance estimation, contaminants and candidate selections, thus we are only interested in analysing how the search design impacts the parameters estimation.

\subsection{Light curve simulations}\label{sec:simulations}

Considering the combinations of KN and GRB events, simulated light curves allow us to build up a science case for KNe which are well localized and that have a constrained estimate for the distance. 
To approach the complexity of the KN models \citep{2020Pang} we use nuclear-multimessenger-astronomy algorithm, \texttt{nmma}\footnote{\url{https://github.com/nuclear-multimessenger-astronomy/nmma}} \citep{2022Pang}. The software gives us the possibility to generate a distribution of realistic ejecta masses descrived by a population of BNS merger,
using the procedure from \citet{Dietrich20}, where they developed a framework to combine multiple constraints on the masses and radii of NSs, including data from GWs, EM observations, and theoretical nuclear physics calculations. 
The simulations to derive the prior distribution of the ejecta mass employ quasi-equilibrium circular initial data in the constant rotational velocity approach, i.e. they are consistent with Einstein equations and in hydrodynamical equilibrium. The model assumes the SFHo \citep[Steiner-Fischer-Hempel baseline model in ][]{2013Steiner} equation of state (EoS), which satisfies the current astrophysical constraints \citep[e.g., ][]{2019Miller}. 

In our study, we employed the KN model introduced by \citet{2017Perego}. This model takes into account the radiation produced by two distinct components: dynamical ejecta and disk ejecta. The disk ejecta can be further divided into two parts. The first part is wind ejecta \citep{1997Ruffert, 2015Kiuchi, Fernandez_2017}, which is propelled in directions close to the polar axis by the neutrino flux originating from the hotter regions of the disk during the neutrino-dominated phase. The second part of the disk ejecta is known as secular ejecta \citep{2013Rodrigo,2018Radice}, arising from viscous angular momentum transport.

By analyzing the distribution of ejecta masses, we were able to derive the distribution of ejecta velocities. This velocity distribution is influenced by the explosion energy, and for more details on this aspect, one can refer to \citet{2017Metzger}.

To take into account all the possible parameter correlations we construct priors distribution for the KN parameters shown in \autoref{fig:apriorimass}. The distribution of the model parameters we injected in \texttt{nmma} to simulate KN light curves, the ejecta mass and velocity distribution is modeled by the binary mass distribution in \citep{Dietrich20},  the open angle for the wind ejecta component is drawn by a uniform distribution $\mathcal{U}(\pi/6,\pi/4)$. Then drawing randomly from these distributions we characterise the entire sample of simulated sources. 

To create a science case to frame the experiment we made some assumptions: 
\begin{itemize}
    \item we have the information of the distance to the event thanks to a GW trigger;
    \item we have the information of the energy of the explosion because we detect a correlated GRB;
    \item if there is an associated GRB we can assume the localization of the KN as known.
\end{itemize}

We combine the simulated population with a single afterglow model for each viewing angle, so that the differences in the resulting light curves are due to the KN contribution (the model parameters for the afterglow and the KN are listed in \autoref{knparams} and \autoref{tab:lsst_grbparams}).  This choice is driven by the possibility to detect KNe\ as flux excess in the afterglow evolution \citep{2020Rossi}, and because the frequency of observed afterglow exceed the rate of KN \citep[$\approx 5$ KNe\ $\mathrm{yr}^{-1}$ versus $\approx 100$ afterglows $\mathrm{yr}^{-1}$, see][as reference for the reported afterglows' rate]{2009LSST} we expect to recognise KN using well sampled afterglow light curves. All sources are simulated assuming three reference distances along the line of sight: 42, 100 and 300 Mpc. Eventually, the effect of the viewing angle cannot be neglected, thus we consider also three reference viewing angles of 0, $\pi/4$ and $\pi/2$ rad.
The particular choice was made so that the ability in retrieving the physical parameters from the light curves would not be influenced by effects on the light curve due to the distance (\eg\ selection effects due to the limiting depth of the survey or Malmquist bias) or viewing angle. 
The lower distance correspond to the distance value of AT2017gfo as reference, while the median and higher distances were set based on considerations about the detectable sGRB rate,
 indeed \citet{2020Dichiara} poited out $N_\mathrm{sGRB}(d_L<200~\mathrm{Mpc})= 1.3^{+1.7}_{-0.8}~ \mathrm{yr}^{-1}$ sGRBs within 200 Mpc are detectable. Scaling this values to higher distances we estimated  $N_\mathrm{sGRB}(d_L<350~\mathrm{Mpc})\approx 1.2~\mathrm{yr}^{-1}$ considering the lower bound of the uncertainty range, hence we set the higher reference distance to $300$ Mpc.
 
\begin{table}
\centering
  \begin{tabular}{ccc}
  \hline
  KN parameter & Unit & Value \\
  \hline
    $D_L$        & Mpc & $[42,100,300]$ \\
    $\theta_w$ & $deg$  & $\mathcal{U}(\frac{\pi}{6},\frac{\pi}{4})$   \\
    $v_\mathrm{ej}$ & c & $PDF(0.1,0.6)$ \\
    $M_{ej,dyn}$ & $M_{\cdot}$ & $PDF(0,0.1)$ \\
    $M_{ej,wind}$ & $M_{\cdot}$ & $PDF(0,0.3)$\\       
  \hline
  \end{tabular}
    \caption{\label{knparams} The KN parameters and their probability density functions (PDF) used to create simulated light curves for the KN components with \texttt{nmma}. The luminosity distance , $D_L$, 
    is needed to generate the model, as well as the angle for the wind ejecta polar emission,$\theta_w$ , the ejecta velocity ,$v_\mathrm{ej}$, the ejecta masses, $M_{ej,dym}$ 
    and $M_{ej,wind}$, the extinction, 
    Ebv, the exponent, $\beta$ of the relation 
    $M_v=M(\frac{v}{v_0})^{\beta}$, where M is the total mass, v is velocity of the mass envelope and 
    $v_0$ is the average minimum velocity of the ejecta. The last equation is used to reproduce the structure of the matter within the moving ejecta 
    \citep[see][ for details on KN model]{2017Perego}. 
    }
    
\end{table}

\subsubsection{Kilonova model}

\texttt{nmma} use fitting formulae based on numerical simulation of the merger and post-merger dynamics to compute the ejecta properties \citep{2018Radice,2020Kruger}, as a function of the binary parameters (namely the component masses and the EoS). The procedure used is presented in \citet{Dietrich20}, where they survey 5000 EoS that provide possible descriptions of the structure of NSs recovering those who reproduce astrophysical constraints, such as NS maximum mass. For more details see the \textbf{Supplementary Material} of the referenced paper. 

We evaluate then the accretion disk mass using the fitting formula from \citet{2020Barbieri}, whose predictions are consistent with both symmetric and asymmetric BNS merger numerical simulations presented in \citet{2018Radice}, \citet{2019Kiuchi}, \citet{2020Bernuzzi}, \citet{2020Vincent}.

The computation is based on a semi-analytical model in which axisymmetry relative to the direction of the binary angular momentum is assumed. The ejecta, assumed to be in homologous expansion, are divided into polar angle bins, and thermal emission at the photosphere of each angular bin along radial rays is computed following \citet{2014Grossman,2015Martin}, taking into account the projection of the photosphere in each bin. See \autoref{knparams} for the reference parameters' distribution for the light curves simulations.

\subsubsection{Afterglow model}
\begin{table}
\centering
\begin{tabular}{ccc}
  \hline
  GRB parameter & Unit & Value \\
  \hline
    $D_L$         & Mpc & $[42,100,300]$ \\
    $\theta_{v}$ & rad  & $[0, \frac{\pi}{4}, \frac{\pi}{2}]$   \\
    $\phi_{c}$ & rad & 0.1 \\
    $\theta_{w}$ & rad  & 0.1 \\
    $E_0$ & erg & $10^{53}$ \\
    $n_0$ & cm$^{-3}$ &  0.1\\
        $p$    &   - & 2.2\\
    $\epsilon_e$    &   - & 0.1 \\
    $\epsilon_B$    & -  & 0.01\\
  \hline
  \end{tabular}
  \caption{The GRB parameters fed to \texttt{\texttt{nmma}} and \texttt{afterglowpy} that were used in to create simulated light curves for the GRBs. 
  The luminosity distance , $D_L$, 
  is needed to generate the model, as well as the viewing angle, $\theta_v$, 
  the half-opening angle, $\phi_c$,
  the outer truncation angle,  $\theta_w$, the isotropic-equivalent energy, $E_0$,
  the circumburst density, $n_0$, 
  the electron energy distribution index, $p$, and the fraction of energy imparted to both the electrons, 
  $\epsilon_e$, and to the magnetic field, 
  $\epsilon_B$, by the shock.} 
  \label{tab:lsst_grbparams}
\end{table}

GRBs associated with gravitational-wave events are, and will likely continue to be, viewed at a larger inclination than GRBs without gravitational wave detections. As demonstrated by the afterglow of GW170817A, this requires an extension of the common GRB afterglow models which typically assume emission from an on-axis top hat jet. We used a Python package \texttt{afterglowpy} \citet{2020Ryan} that characterize the afterglows arising from structured jets, providing a framework covering both successful and choked jets. The temporal slope before the jet break is found to be a simple function of the ratio between the viewing angle and effective opening angle of the jet. 

To accommodate an initial structure profile $E(\theta)$ \texttt{afterglowpy} we consider the flux as a function of the polar angle $\theta$. It assumes each constant-$\theta$ annulus evolves independently, as an equivalent top hat of initial width $\theta_j = \theta$. This is a very good approximation when transverse velocities are low: when the jet is ultra-relativistic and has not begun to spread and when the jet is non-relativistic and the spreading has ceased \citet{van_Eerten_2010}. The model allows for several angular structures of the GRB jet.
In our exercise we use one GRB model for the afterglow since we are interested in the ability to infer the KN parameters, the parameters used for the GRB model are shown in \autoref{tab:lsst_grbparams}, and we assumed a Gaussian jet structure to not correlate effects on the viewing angle or beam direction with effects coming from the peculiar jet-environment interaction due to particular geometry. The parameters are set to produce simulations as realistic as possible, for this reason we assumed the parameters from the usual values from both observed and modelled Afterglow population as expressed in \citet{2010Eerten, Davanzo2014, 2018Eerten}.
However specific, the parameters selected for the reference afterglow template are the median value of an observed short-GRB population take as reference from \citet{Davanzo2014}. Thus the reference afterglow template can be considered representative of the short-GRB population. We also considered different reference values for short-GRB from \citep{2015Fong}, however different assumptions appear not to dramatically impact the results, thus we considered the values in \autoref{tab:lsst_grbparams}.

\begin{figure*}[h!]
    \centering
    \hspace*{-3.0cm}
    \includegraphics[scale=0.3]{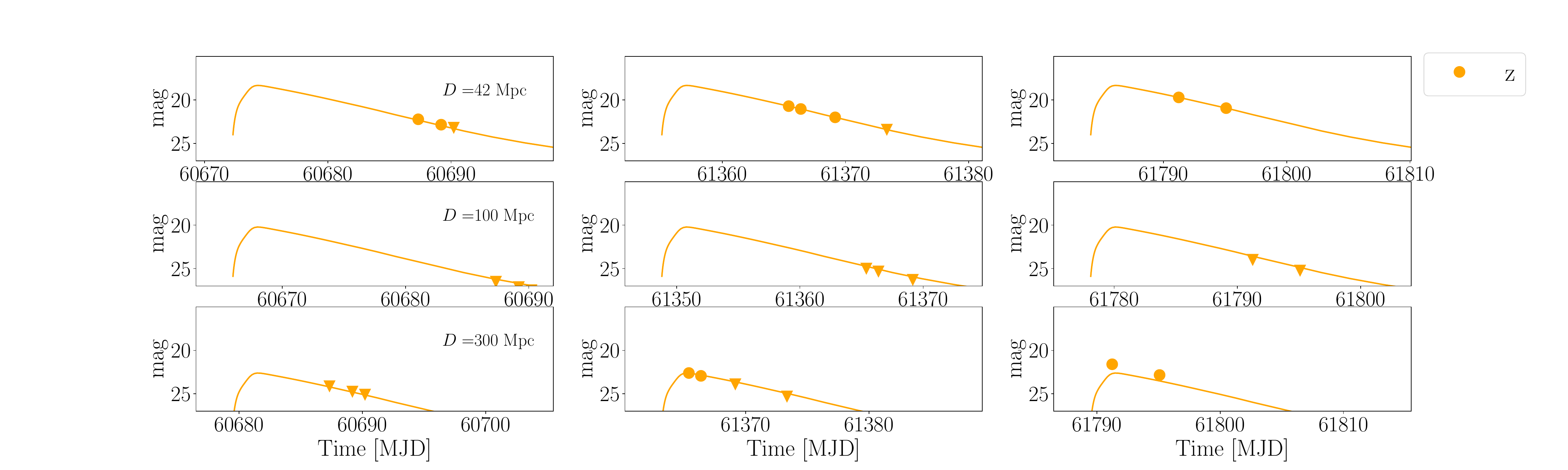}
    \caption{Here an example of a single template (tick lines) observed at $(RA,DEC)=(197.45 ,-23.38)$ (the position of the NGC4993, the host galaxy of AT2017gfo), at the three reference distances $[42, 100, 300]$ Mpc during three time-windows through the 10 years of the survey. To reproduce the observed detection showed in the panels as filled colored points we used the \texttt{baseline\_ v2.0\_10yrs} strategy design. The triangles represent no-detections. The way the strategy plans to look at the footprint imply that we will be able to detect the events but 
    we won't have the same ability to characterize as we will lose information about the color and the morphology of the light curve. Indeed, we simulate the light curve in all the bands but only z-band appears to be detectable in this region and survey time. }
    \label{fig:lc_obs}
\end{figure*}

\section{MAF and \opsim~}\label{sec:MAF}

The comprehensive discussion of the software made available by Rubin Observatory for community contribution to the survey design is not within the scope of this paper. Interested readers are referred to the opening paper \citet[][hereafter B21]{frontpaper} and its references for a full examination of the software's workings \citep{2019AJ....157..151N, 2016SPIE.9910E..1AY, opsim, 2014Delgado} for more details and information on this topic.

The Operations-Simulator software (\opsim\footnote{ \url{https://www.lsst.org/scientists/simulations/opsim}}) \citep{2014Delgado}
 generates a simulated strategy based on a set of criteria, such as total number of images per field per filter, including simulated weather, telescope downtimes and other occasional interruptions. The survey requirements (survey strategy) are the input to an \opsim~run and the output is a database of observations with associated attributes (e.g. image $5\sigma$ depth) that specify a succession of simulated observations for the 10-year survey. Since its creation, the Rubin \opsim~has gone through various revisions, the main difference of which are the methods used to optimize the pointing sequences and filters to achieve the desired survey features (\citetalias{frontpaper}).

The Metric Analysis Framework (\maf\footnote{ For more informations about the \maf~ you can refer to \url{https://www.lsst.org/scientists/simulations/maf}}) API is a software package created by the Rubin Observatory \citep{maf} to evaluate how various simulated LSST (Large Synoptics Survey Telescope) observation strategies impact different specific science goals. 
The \maf~has been made public upon its creation to facilitate community input in the strategy design and it enables interaction with \opsim~primarily by \texttt{SQL}, allowing the user to select filters or time ranges (\eg~the first year of the survey). Further, the choice of \texttt{slicers} allows the user to group observations. For example, one may ``slice'' the survey by equal-area spatial regions, using the HEALPIX scheme of \citealt{healpix2005}. Throughout, we choose a \texttt{HealpixelSlicer} with resolution parameter \texttt{NSIDE}=16, corresponding to a pixel area of 13.4 square degrees  \citep[and thus the choice that most closely matches the size of the Rubin LSST field of view, see ][ for reference to Rubin LSST characteristics]{2019Ivezi}.
Thus, to pass from the simulated theoretical lightcurves produce according the procedure described in \autoref{sec:method} to the simulated observed lightcurves we used the \maf. This way we are able to apply parameter estimation tools to the simulated observation to analyse the impact of the observing strategies on the ability to retrieve in parameters injected in \texttt{nmma} to produce the theoretical simulations (more details in \autoref{sec:opsim}. 

\section{Observing strategy impacts on parameter estimation}\label{sec:performance}
\begin{figure*}
  \centering

  \begin{minipage}[b]{0.5\linewidth}
    \centering
    \includegraphics[width=\linewidth]{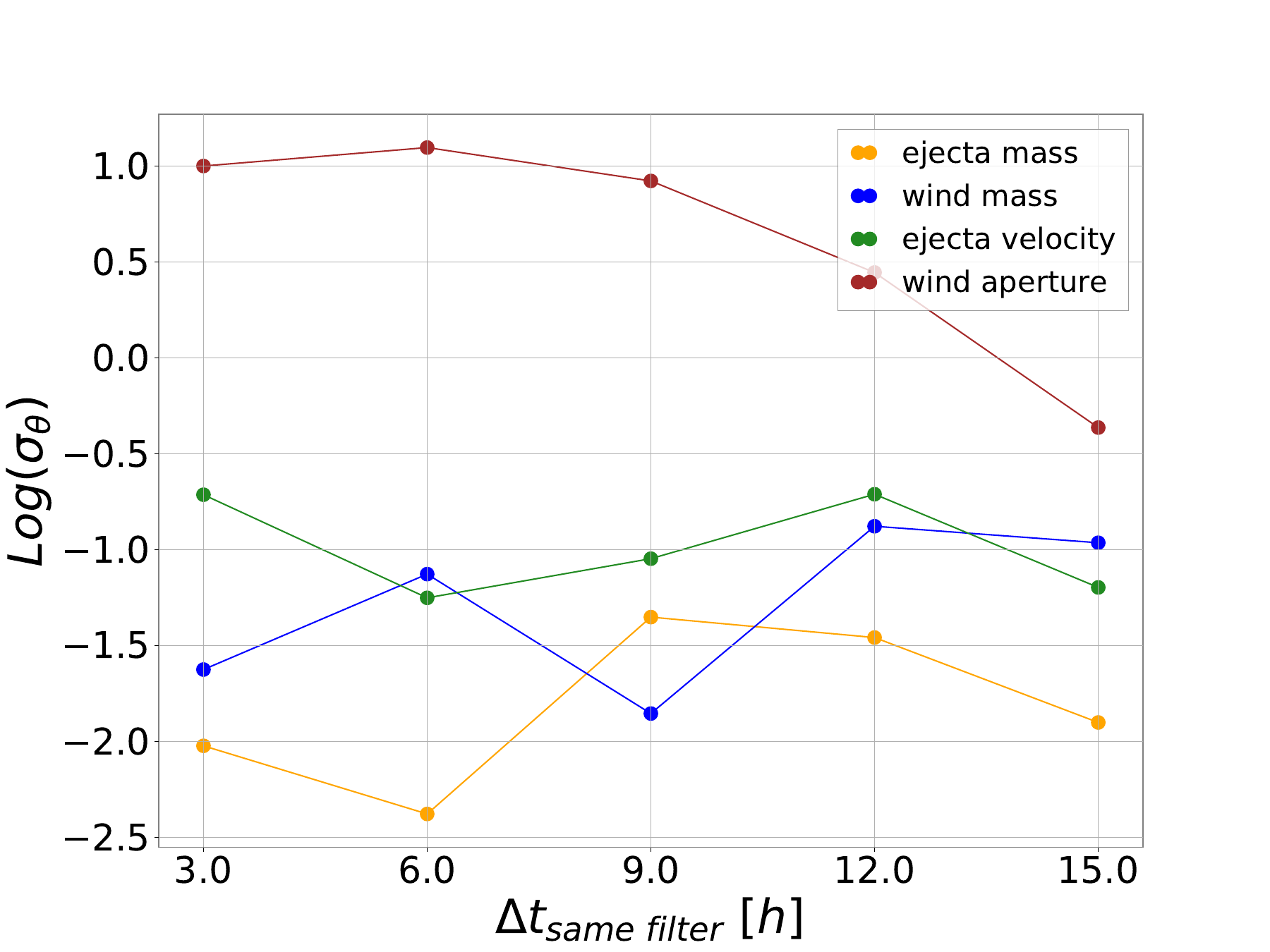}
  \end{minipage}
  \hfill
  \begin{minipage}[b]{0.45\linewidth}
    \centering
    \includegraphics[width=\linewidth]{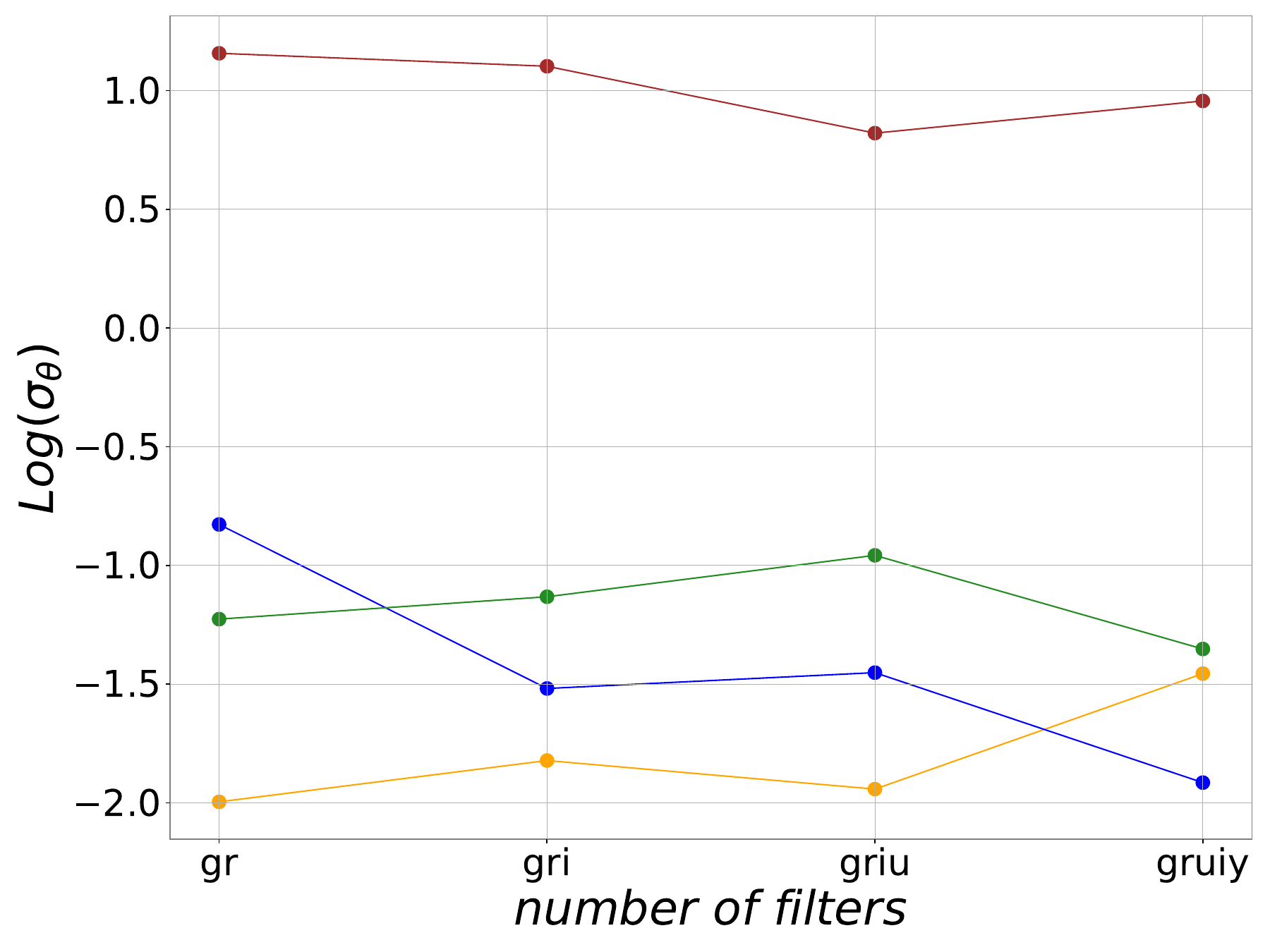}
    
  \end{minipage}

  \vspace{0.5cm} 

  \begin{minipage}[b]{0.5\linewidth}
    \centering
    \includegraphics[width=\linewidth]{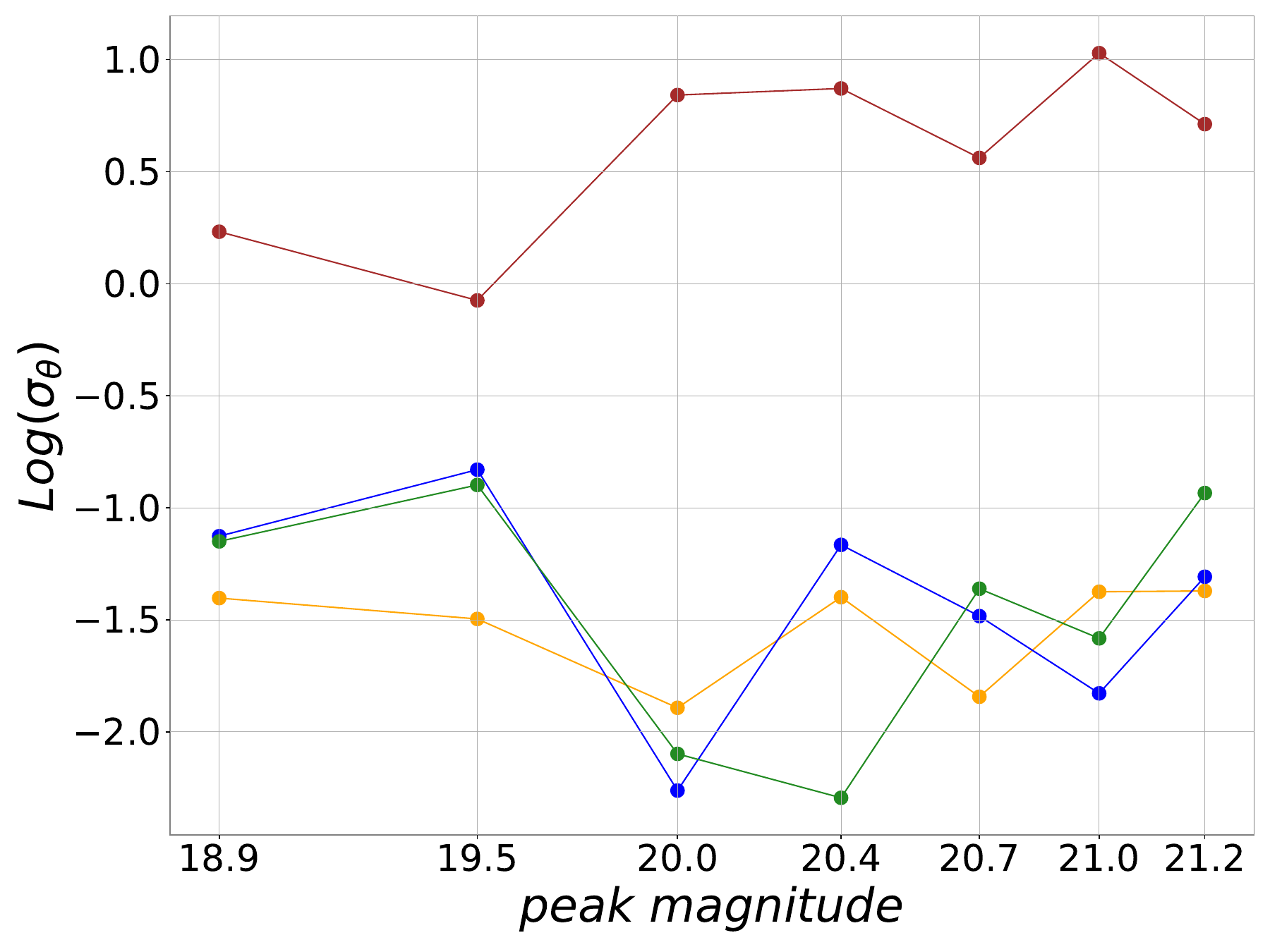}
    
  \end{minipage}

  \caption{The fitting performance analysis for a single light curve. Each panel shows the logarithm of the posterior's variance as a function of the time gap between two consecutive points on the light curve (upper left panel); the number of available filters (upper right panel); the light curve maximum magnitude (bottom panel). The plot can be read as follows: lower values on the y-axis represent better performances. Interpreting the plots, we find that the most information is gained when we populate the light curve with very close points in time, and the event is bright (panels upper left and bottom panels). There is a net improvement in the performance when considering all the filters together; however, in the other cases, the difference in performance is negligible.}
  \label{fig:cornersystematic}
\end{figure*}

\begin{figure*}
  \centering
  \includegraphics[width=1.\linewidth]{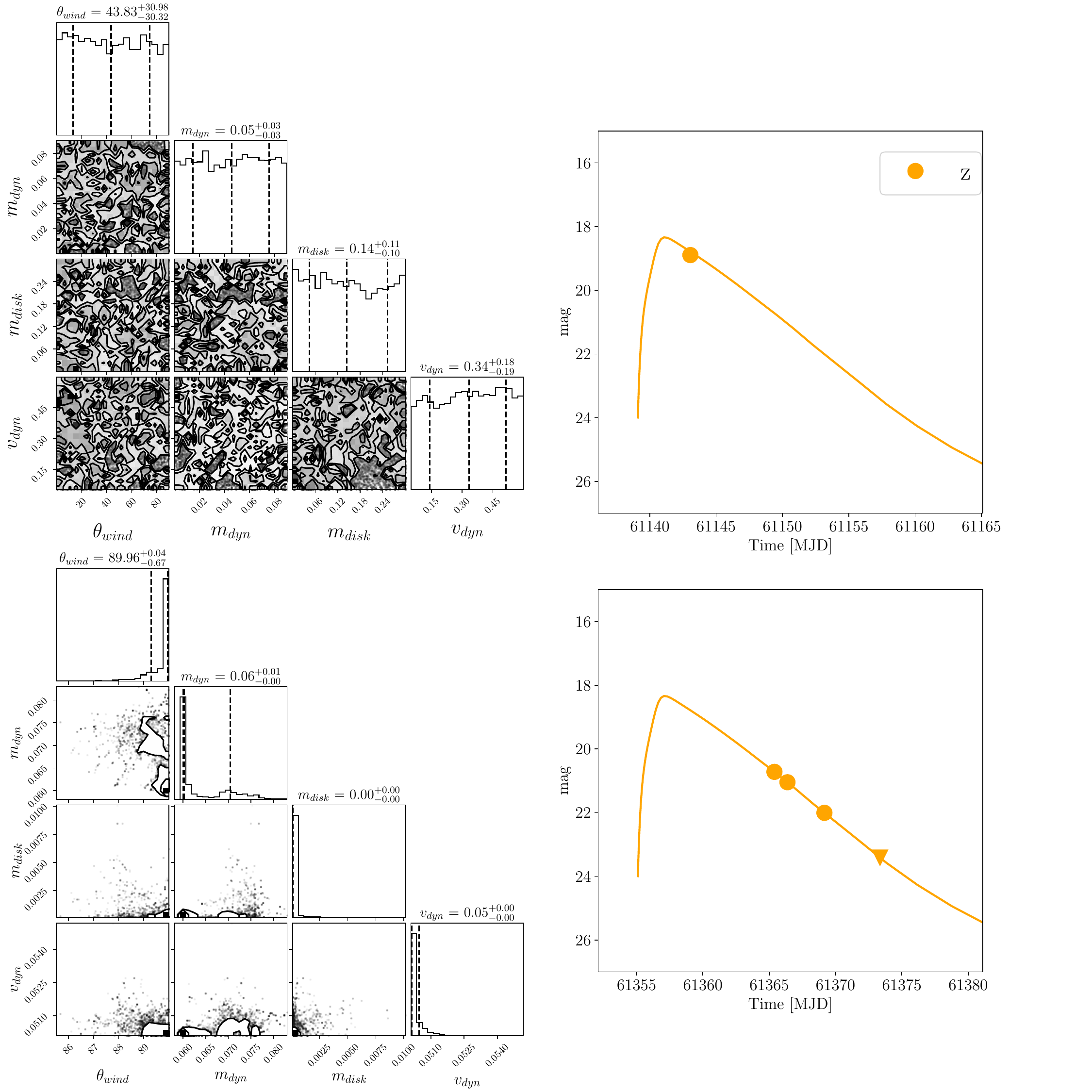}
  \caption{Corner plots with an example for a worst case scenario (top row) and best case scenario (bottom row) when applying the fitting procedure.}
  \label{fig:fitexamples}
\end{figure*}

The ability to populate light curves is very limited, as the number of filters and the number of detections in each filter vary depending on the intrinsic properties of the event. This could impact our capability to infer the KN parameters.
To evaluate the performance of \texttt{\texttt{nmma}} in estimating the KN parameters and reproduce the injected light curve, we used as metric for the performance of the fitting procedure the posterior's variance.
We analyze three features that typically impact a sampler performance:
\begin{itemize}
    \item the number of detected points on the light curve;
    \item the number of available filters;
    \item the light curve peak magnitude.
\end{itemize}

\new{In the upper left panel of \autoref{fig:cornersystematic}  we analyzed the light curve sampling by changing the time resolution of the template. 

When we refer to the general trend, the figure shows that for light curve data spaced more than $\approx 6h$, dynamical ejecta is more poorly constrained with respect the other parameters, 
with the exception of two cases. However in those two cases the value of ejecta mass and velocity are closer to the others, suggesting that the effect could be related to small fluctuation around the best fit parameter configuration.
This is interpreted as an effect of the fitting procedure, because of the time gap between the detections we miss the possibility to get the maximum, however constraining the rise and the fall of the light curve would help in constraining the KN model parameters.}
Specifically, the sampler cannot constrain the global minimum of the cost function (in our case the likelihood of the detections, see \autoref{sampler}) in the $M_\mathrm{dyn}$,$v_{dyn}$,$M_\mathrm{disk}$, $\theta_{w}$ hypercube when the light curve is not populated. This is because there is a large degeneracy of states which reproduce the same collection of fluxes (bottom panel in \autoref{fig:fitexamples}). 
When the sampler can constrain in the hypercube the values of the parameters the performance appear to be higher (upper panel in \autoref{fig:fitexamples}). 
This happen for the other panels too, however when analysing the performance of the sampler with respect the number of filters (middle panel in \autoref{fig:cornersystematic})
we see this is not a very important observational feature to constrain the posterior's variance. For the experiment show in 
 middle panel of \autoref{fig:cornersystematic} we set the detection cadence at 9 hours.  

Eventually the peak magnitudes impact similarly to the upper left panel on the performance as shown in \autoref{fig:cornersystematic} bottom panel. \new{Thus, this behaviour can be interpreted similarly to the case showed in the upper left panel, considering that the closer to the limiting magnitude the peak is, less detections on the light curve features we get.}
The main problem is that the degeneracy in the parameter space produces different minimum in the cost function because the systematics we analyzed here. Thus, the area surveyed to find the global minimum is higher. Eventually we are able to infer the values of the injected parameter within some confident level with the drawback to lose precision. 

\section{Model's description impacts on parameters' estimation}\label{sec:model_impact}

\begin{table*}

    \hspace*{-1.3cm}
    %\begin{adjustbox}{scale=0.8}
    \fontsize{18}{16}\selectfont
    \resizebox{\textwidth}{!}{\begin{tabular}{cc}\hline
        &$\boldsymbol{A}_q^{m_\mathrm{ej}}(M_1)$\\ \hline
        
        $q<1$ & 
        \begin{tabular}{c}
            $\left\{\left[\frac{\alpha_{m_\mathrm{ej}}}{M}q^{-\frac{2}{3}}(2C_1q-3)-\beta_{m_\mathrm{ej}}nq^{-(n+1)}-\frac{\beta_{m_\mathrm{ej}}n}{M_1}q^{n-2}\right.\right.$\\
            $\left.-\frac{1-2C_1q}{M_1C_1}q^{-\frac{8}{3}}-\left(\frac{1-2C_1}{C_1}\right)q^{-\frac{4}{3}}\right]^2+$\\
            $\left[\left(\frac{1-2C_1q}{C_1}\right)\frac{\alpha_{m_\mathrm{ej}}}{3}q^{-\frac{5}{3}}-\beta_{m_\mathrm{ej}}nq^{n-1}-\frac{\beta_{m_\mathrm{ej}}n}{M_1}q^{-(n+2)}\right.$\\
            $\left.\left.+\left(\frac{\alpha_{m_\mathrm{ej}}}{M_1}(2C_1-3)-\frac{\alpha_{m_\mathrm{ej}}}{3}\frac{a-2C_1}{M_1C_1}\right)q^{-\frac{7}{3}}\frac{\alpha_{m_\mathrm{ej}}}{M_1}(2C_1-3)q^{-\frac{1}{3}}\right]^2 + \frac{\operatorname{sech}^4\left(\frac{\hat{\Lambda}-\gamma_{disk}}{\delta_{disk}}\right)}{\delta^2_{disk}}\right\} \frac{q^2}{1-2q}$
        \end{tabular}
        \\ \hline
        $q\approx1$ &
        
        \begin{tabular}{c}
            $\left\{\left[\frac{\alpha_{m_\mathrm{ej}}}{3}\left(\frac{1-2C_1}{C_1}\right)+\beta_{m_\mathrm{ej}}n\left(1-\frac{1}{M_1}\right)\right]^2+\right.$\\
            $\left.+\left[\beta_{m_\mathrm{ej}}+\frac{\alpha_{m_\mathrm{ej}}}{M}(2C_1-3)+\alpha_{m_\mathrm{ej}}\left(\frac{1-2C_1}{C_1}\right)\left(1-\frac{\alpha_{m_\mathrm{ej}}}{M_1}\right)\right]^2 \right.+ $\\ $\left. \frac{sech^4\left(\frac{\hat{\Lambda}-\gamma_{disk}}{\delta_{disk}}\right)}{\delta^2_{disk}}\right\}\frac{2q^2}{1-2q}$
        \end{tabular}
        
       \\ \hline \hline
         &$\boldsymbol{A}_q^{v_\mathrm{ej}}(M_1)$\\ \hline
        $q<1$ & $\frac{1}{q^4}\left[5M_1 \left(\frac{\alpha_{v_\mathrm{ej}}\gamma_{v_\mathrm{ej}}G}{c^2R_1}\right)^2+\alpha_{v_\mathrm{ej}}\left(\frac{4}{M_1}+2\right)\right]$\\ \hline
        $q\approx1$  &  $M_1\left(\frac{\alpha_{v_\mathrm{ej}}\gamma_{v_\mathrm{ej}}G}{c^2R_1}\right)^2$\\ \hline
    \end{tabular}}
    %\end{adjustbox}
    \caption{A relation between these variable can be found in \citet{1996Timmes} : $M_i-M_i^{\ast} = 0.075\times (M_i^{\ast})^2$. 
    For a NS the assumption $M_i-M_i^{\ast} \approx 0$ has been considered. The tidal deformability parameter,$\hat{\Lambda}$ is modeled from \citep{2020Chatziionnau}.} 
    \label{tab:unctheory}

\end{table*}

\begin{figure*}
    \gridline{\hspace*{-1cm}
    \fig{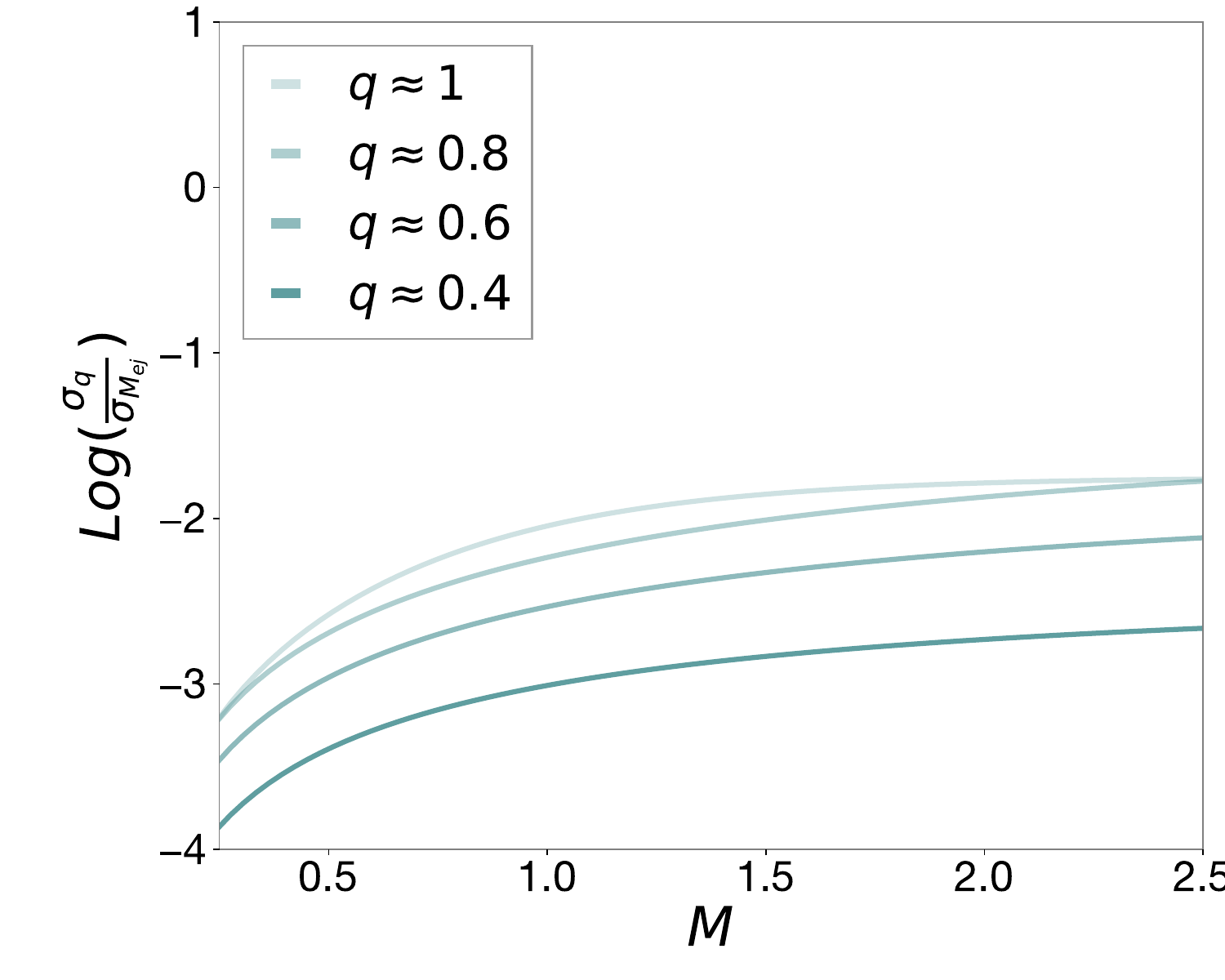}{0.5\textwidth}{($a$)}
    \fig{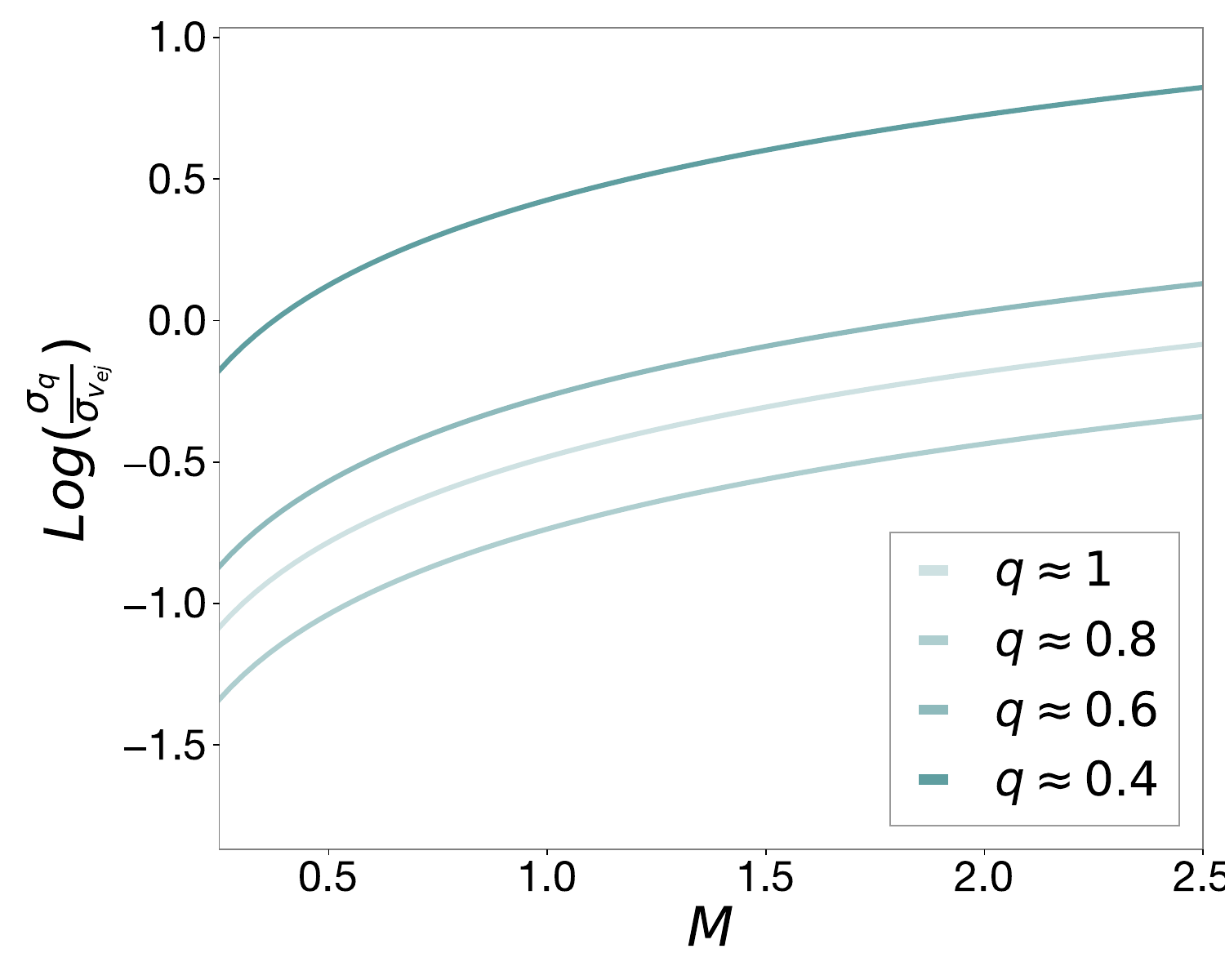}{0.5\textwidth}{($b$)}}
    \caption{The sensitivity plot described in detail in Section \ref{sec:model_unc}, 
    the panels show the parameters sensitivity to the variation of the BNS represented by the primary mass, and the mass ratio, q.}
    \label{fig:fig_unctheory}
\end{figure*}
\label{sec:model_unc}

The assumptions on the radiation transport and on the nuclear network, at the foundation of a model that attempt to describe the observed event, can influence the ability of the cost function to find the global minimum of the parameter space that gives the best parameters' configuration to describe them. 
Due to this connection between the model and the cost function, we analyze if the behavior we highlighted in the previous paragraph and in \autoref{fig:fig_unctheory} is somehow related to the model used in the fitting procedure.

We consider a BNS population from \citet{Dietrich20} as pointed out in \autoref{sec:method} and fit to currently available observational constraints from both GW-detected and Galactic BNS binaries as described in Appendix A in \citet{2022Colombo}. The merger rate is obtained by convolving the delay time distribution (represented as the time gap between the formation of the binary system and its merger) with the cosmic star formation rate \citet{2014Madau} normalized to the local rate density $R_0= 347^{+536}_{-256}\mathrm{Gpc}^{-3}\mathrm{yr}^{-1}$ in order to reproduce the number of significant BNS events \citep{2021AbbottLIGO3}.

For each event, the signal to noise ratio (SNR) of the GW strain has been evaluated for the LVK detectors. For events above the SNR threshold in the population, $M_\mathrm{ej},~v_\mathrm{ej}$ are estimated using equations (18) and (22) in \citet{2018Radice} and considering the SFHo EoS:

Taking into account that both dynamical and disk ejecta contribute to the mass of the ejecta, the total $m_\mathrm{ej}$ is estimated as $m_\mathrm{ej}= M_\mathrm{dyn}+M_\mathrm{disk}$. 
We aim to analyze the impact of observations on constraining the mass ratio, thus with $X=[m_\mathrm{ej},v_\mathrm{ej}]$ we estimated the uncertainties on the KN parameters:
\begin{align}
\sigma_{X}^2 &=\left| \frac{\partial X}{\partial M_1}\right|^2\sigma^2_{M_{1}} +\left| \frac{\partial X}{\partial M_2} \right|^2\sigma^2_{M_{2}} =\\
&=\left[ \left| \frac{\partial X}{\partial M_1}\right|^2 +\left| \frac{\partial X}{\partial M_2} \right|^2\right]\sigma^2_{M_1},&\nonumber
\end{align}

where we consider the uncertainties on the two merging NS masses to be equal in the last equation.
Assuming the uncertainties on the masses as,

\begin{equation}
\sigma_q^2=\frac{q^2(1+q^2)}{M_1}\sigma_{M_1}^2.
\end{equation}

Finally, we obtain the following,

\begin{align}
\sigma_{X}^2 &=\left[ \left| \frac{\partial X}{\partial M_1}\right|^2 +\left| \frac{\partial X}{\partial M_2} \right|^2\right]\frac{M_1}{q^2(1+q^2)}\sigma^2_{q}=&\\
&=\boldsymbol{A}^X_q(M_1)\sigma^2_{q},&\nonumber
\end{align}

where $\boldsymbol{A}^X_q(M_1)$ -- which is the ratio between the variance squared of the mass ratio and the ejecta mass or velocity -- can be interpreted as the sensitivity to the system's photometric observation, in \autoref{tab:unctheory} are shown the form for $\boldsymbol{A}^X_q(M_1)$ for $q<0~q>0~\text{and}~q\approx 1$ related to ejecta mass or ejecta velocity estimations (see \autoref{fig:fig_unctheory}.

The inference of model parameters from a physical model assumes that the value inferred represents the value from the event's underlying model. However due to over-simplifications in the theoretical treatment or technological limitation this is not always true. 
\autoref{fig:fig_unctheory} shows how the model impacts on the parameters values (i.e. ejecta mass and ejecta velocity) when we try to infer them assuming we can measure other observables: the total binary mass and the mass ratio. 
If the physical model used to describe the event is highly degenerate for those parameters, we cannot distinguish between a set of parameters that produce the same set of observables, i.e. the uncertainties on the parameters are so high that the range of possible inferences related to that measure is very broad. 
Our case shows this is the case when we infer the ejecta velocity. Indeed, \autoref{fig:fig_unctheory} panel (b) shows that if we survey the parameter space following the $v_\mathrm{ej}$ the fact that we have high uncertainty translate in a broader region of local minima in the cost function. Thus every inference we make to look for the global minimum we are likely to end up in a very similar state to what we started; this is because the uncertainty distribution is almost uniform, which means that whatever the true ejecta velocity is the chance of being in any other region close to that value is the same, meaning that we are likely to miss the global minimum.
Conversely, \autoref{fig:fig_unctheory} panel (a) shows the case in which we survey the parameter space in the direction of the ejecta mass $m_\mathrm{ej}$. This direction of the parameter space appears to be very helpful in constraining the specific parameter, showing that the uncertainty on the inferred parameter can change by an order of magnitude. Hance, the possibility of matching the global minimum can be higher if we survey the $m_\mathrm{ej}$ direction of the hypercube.
Eventually, we can conclude that we expect to have very well constrained inference for ejecta mass with respect ejecta velocity.

\section{Observing constraints from the simulations}\label{sec:opsim}

\begin{figure*}[b!]
    \centering
    \includegraphics[scale=0.35]{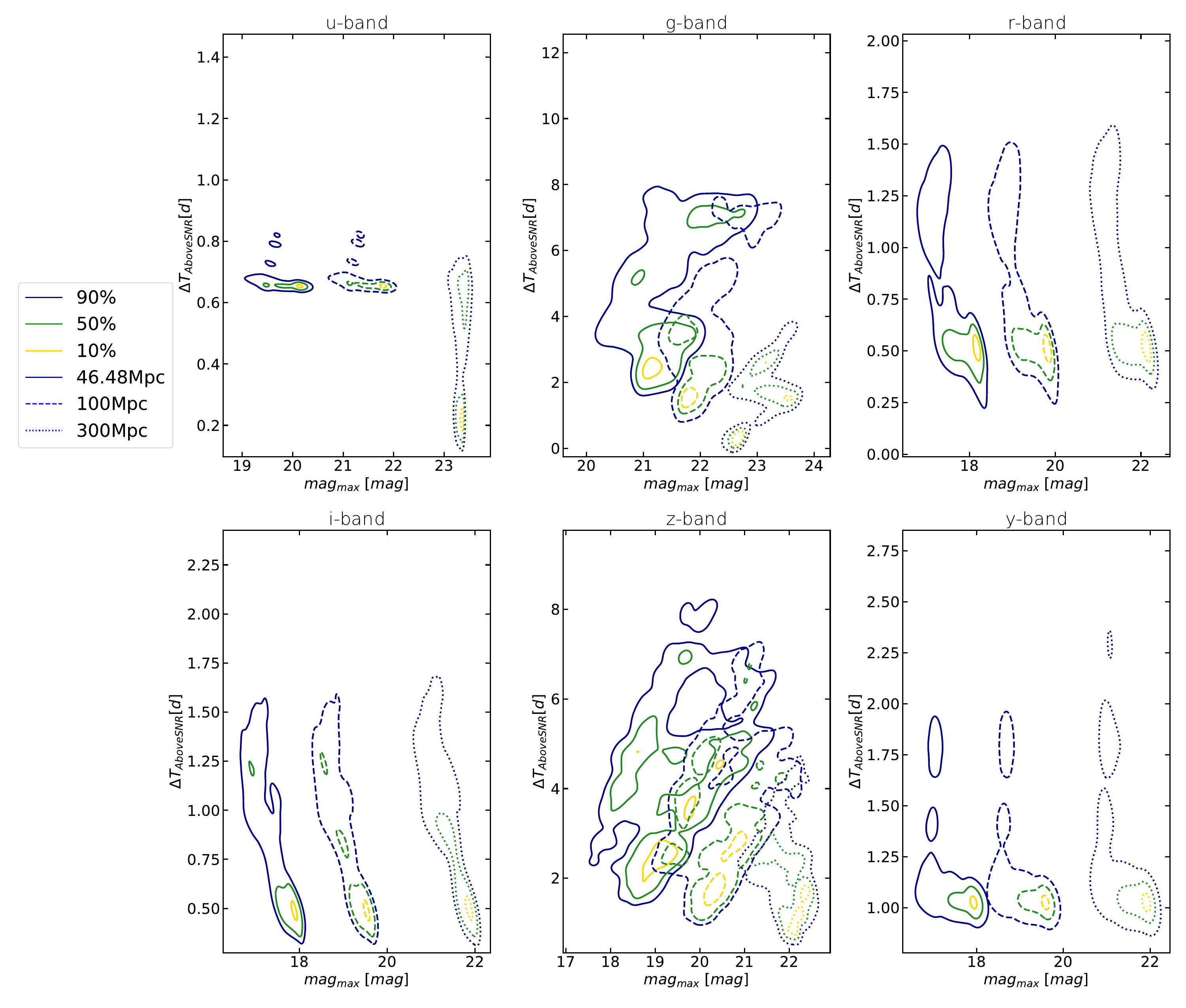}
    \caption{The distribution of the features we can extract from the observed light curves. 
Each panel represents the distribution of two features: peak magnitude and duration of the light curve above the limiting magnitude; 
    different lines represent a distance at which the event is simulated. Details in \autoref{sec:opsim}}
    %Each column represents the distribution of a %feature: peak magnitude and duration of the light %curve above the limiting magnitude; 
    %and each row represents a distance at which the %event is simulated. Details in %\autoref{sec:performance}.}
    \label{fig:lc_features}
\end{figure*}

\new{In section \autoref{sec:method} we described how we simulated light curves, with a combination of KN and afterglow emissions from the same source. These simulations are then used as reference templates to produce mock observed KN light curves during the operation time of LSST. }
To tackle the problem, we associate to the KN + Afterglow light curves templates an explosion time, uniformly chosen, within the 10 yr of the survey all over the observed sky.
Eventually, the ability to discover fast and faint transients, such those we simulated, largely depends on the area observed -- which in our experiment is the field of view of the pointings --, the depth of those observations, the cadence, and the filters adopted by the survey. 

Using the \texttt{baseline} as a test for our machinery, we applied the observational constraints from this \opsim~ to simulate what the observed light curves look like. We simulated for each reference distance and viewing angle a set of 100000 KNe + Afterglows events, with a total of 900000 sources. 

\autoref{fig:lc_features} shows the contours of the detectability regions change dramatically for further events, showing that the best filters to follow up KN+Afterglow events are g + z bands, due to the possibility to follow the events for more time up to 100 Mpc. 
Whereas optical filters show to be very performative on the bluer side of the spectrum with closer events, the relative importance between bluer and infrared filters appear to be unchanged as the source's distance increases.
Analysing the detected events -defined as all the events which light curve have more then two detected observation in any filter- we find that as the source is more distant, events will be observable for a the time range almost equal the time gap between two consecutive observations. Light curves with this duration above the SNR will have just two detected point in the specific wavelength, thus a multi wavelength analysis is mandatory to be able of using these data otherwise no parameter estimation will be possible. \autoref{fig:lc_features} does not change dramatically under the assumption of different viewing angles, thus for all the cases this figure appear to be a good reference for the description of the results.

%From the bottom panel of left column in \autoref{fig:lc_features} can be noted %that the relative number of detected events is higher in these bands even %though the ones in blues bands are much fainter. Whereas optical filters show %to be very performative on the bluer side of the spectrum with closer events, %the relative importance between bluer and infrared filters appear to be unchanged as the source's distance increases.
Because of the nominal limiting magnitudes \citep{2009LSST} we have a very small time window to catch farther KNe\, with a maximum duration of $\sim 6$ days in NIR bands for the simulated sources.
The non-uniformity of the filter coverage through the entire survey impacts on the possibility of characterizing the explosions. Moreover, even though simulations are produced in all the six LSST bands $ugrizy$,  light curves simulated in \autoref{fig:lc_obs} show only the filter that produced a detection for the specific event in that time window $[60000-62000 ~MJD]$.  
Late time evolution of the light curve 
will not be detectable for sources close to 300 Mpc, thus, to be able constraining model parameters a higher priority would rather be given to closer sources if and when detected, because they will be characterized by a well populated light curve. 
The drawback that has to be stressed is the very small time window in which the light curve is detectable, which is for closer sources (\ie~ $42$ Mpc in our simulations) from 5 to 10 days and for farther sources (\ie~ $300$ Mpc in our simulations) from $\leq 1$ to 5 days from the explosion.
For simulated event at $300$ Mpc, the main features that are affected are the peak magnitude and the duration of the event above the limiting magnitude, this is because we get less flux from farther events and thus we reach earlier the limiting magnitude of the survey when observing the light curve's evolution.
Transversely, this impact on the fall rate distribution, which cannot be accurately measured in all the cases because we lose information of the late time morphology of the light curve. 

\section{Constraints on kilonova model}\label{sec:param_estim}
\begin{figure*}
    \centering
    \includegraphics[scale=0.4]{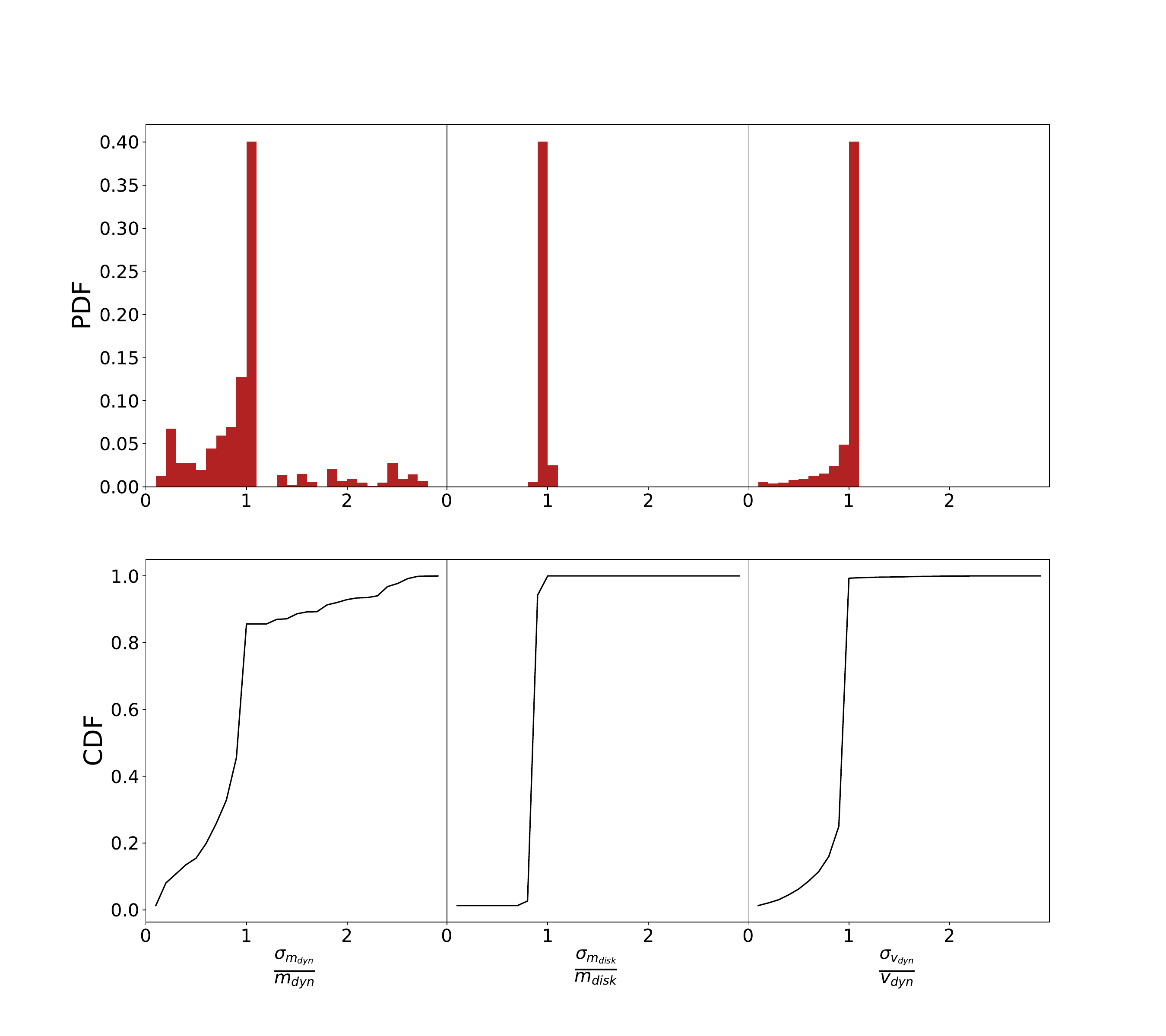}
    \caption{The uncertainties distribution of the model parameters from the fit. Top panels represent PDF of the parameters' relative error; bottom panels represent the cumulative density function (CDF) of the same variable.}
    \label{fig:fit_unc}
\end{figure*}
\new{Follow up of the events in one or more filters will allow us to infer within the $3\sigma$ uncertainties the model's parameters.}
We perform the fit using \texttt{dynesty} \citet{sergey_koposov_2022_7215695}, Python package to estimate Bayesian posteriors and evidences (marginal likelihoods) using Dynamic Nested Sampling methods.
By adaptively allocating samples based on posterior structure, Dynamic Nested Sampling has the benefits of Markov Chain Monte Carlo algorithms that focus exclusively
on posterior estimation while retaining Nested Sampling’s ability to estimate evidence and sample from complex, multi-modal distributions. Nested sampling is a method for estimating Bayesian evidence that was first proposed and developed by \citep{Skilling}. The basic idea is to approximate the evidence by integrating the prior in nested  
 ``shells'' of constant likelihood. Unlike Markov Chain Monte Carlo (MCMC) methods, which can only generate samples proportional to the posterior, Nested Sampling simultaneously estimates both the evidence and the posterior (see \autoref{sampler}). 
\textbf{\new{\autoref{fig:fig_unctheory} shows that the model itself acts as a source of uncertainty on the estimation of KN parameters, with grater uncertainties ratio on the ejecta mass and velocity as the total mass of the progenitor system grows, the behaviour of the curve suggests that the uncertainties of the parameters tend to be lower when the system is more massive}. However when we consider a fixed mass for the progenitor system we measure smaller uncertainties on the ejecta mass as the mass ratio increase, while we note an opposite behaviour for the ejecta velocity. The trend of the uncertainties with respect the total mass of the system implies that there is high degeneracy on the value of the parameters, when looking for the best configuration to replicate the observed light curves. This because the model reproduce similar light curves for all the combination of parameters within the uncertainties' range. }
\autoref{fig:fit_unc}, indeed, shows for the majority of the events, the uncertainty on all parameters is $100 \%$, however, for simulated events with higher ejecta mass and smaller velocities, there is a tail of $\approx 10\%$ events with better constrained parameters. This is important because to constrain the EoS models we need very accurate measurement of KN parameters, so the $30\%$ accuracy, even if good, it is not sufficient, thus the need of a dedicated target of opportunity (ToO) appears to be a necessity to KNe\ follow up from external trigger if the baseline will be the chosen strategy for the survey.

\begin{longtable*}{ll|ll|ll}
\caption{Opsim's Names and IDs, as plotted in Figs. \ref{fig:opsimcomparison_single} and \ref{fig:opsimcomparison}}\label{table:opsim_names} \\
\hline
ID    & Opsim name    & ID    & Opsim name & ID    & Opsim name                              \\
\hline
%\hline
%\scriptsize
0   & baseline & 42 & too\_rate10 & 84 & retro\_baseline  \\
1   & bluer\_indx0  & 43 & north\_stripe & 85  & rolling\_ns2\_rw0\_5  \\
2   & bluer\_indx1  & 44  & twilight\_neo\_nightpattern7v2\_0\_10yrs\_db  & 86 & rolling\_ns2\_rw0\_9  \\
3   & ddf\_frac\_ddf\_per0\_6  & 45  & twilight\_neo\_nightpattern6v2\_0\_10yrs\_db & 87  & rolling\_ns3\_rw0\_5  \\
4   & ddf\_frac\_ddf\_per1\_6  &  46  & twilight\_neo\_nightpattern1v2\_0\_10yrs\_db  & 88  & rolling\_ns3\_rw0\_9  \\
5   & long\_gaps\_nightsoff0\_delayed1827  &  47  & twilight\_neo\_nightpattern2v2\_0\_10yrs\_db  & 89  & rolling\_all\_sky\_ns2\_rw0\_9  \\
6 & long\_gaps\_nightsoff0\_delayed-1  &  48  & twilight\_neo\_nightpattern4v2\_0\_10yrs\_db  & 90  & rolling\_bulge\_ns2\_rw0\_5  \\
7   & long\_gaps\_nightsoff1\_delayed1827  &  49  & twilight\_neo\_nightpattern5v2\_0\_10yrs\_db  & 91  & rolling\_bulge\_ns2\_rw0\_8   \\
8   & long\_gaps\_nightsoff1\_delayed-1  &  50  & twilight\_neo\_nightpattern3v2\_0\_10yrs\_db  & 92  & rolling\_bulge\_ns2\_rw0\_9  \\
9   & long\_gaps\_nightsoff2\_delayed1827 &  51  & local\_gal\_bindx1  & 93  & rolling\_bulge\_6  \\
10  & long\_gaps\_nightsoff2\_delayed-1  &  52  & local\_gal\_bindx2  & 94  & roll\_early  \\
11  & long\_gaps\_nightsoff3\_delayed1827 & 53  & local\_gal\_bindx0 & 95  & six\_rolling\_ns6\_rw0\_5  \\
12  & long\_gaps\_nightsoff3\_delayed-1  & 54  & carina  & 96  & six\_rolling\_ns6\_rw0\_9 \\
13  & long\_gaps\_nightsoff4\_delayed1827 & 55  & short\_exp &  97  & vary\_expt  \\
14  & long\_gaps\_nightsoff4\_delayed-1  & 56  & smc\_movie & 98  & vary\_gp\_gpfrac0\_01  \\
15  & long\_gaps\_nightsoff5\_delayed1827 & 57 & multi\_short & 99 & vary\_gp\_gpfrac0\_05 \\
16  & long\_gaps\_nightsoff5\_delayed-1 & 58  & noroll & 100 & vary\_gp\_gpfrac0\_10   \\
17  & long\_gaps\_nightsoff6\_delayed1827 & 59  & presto\_gap1\_5\_mix  & 101 & vary\_gp\_gpfrac0\_15  \\
18  & long\_gaps\_nightsoff6\_delayed-1 & 60 & presto\_gap1\_5 & 102 & vary\_gp\_gpfrac0\_20 \\
19  & long\_gaps\_nightsoff7\_delayed1827 & 61  & presto\_gap2\_0\_mix & 103 & vary\_gp\_gpfrac0\_25 \\
20  & long\_gaps\_nightsoff7\_delayed-1 & 62 & presto\_gap2\_0 & 104 & vary\_gp\_gpfrac0\_30 \\
21  & long\_gaps\_np\_nightsoff0\_delayed1827  & 63  & presto\_gap2\_5\_mix & 105 & vary\_gp\_gpfrac0\_35 \\
22  & long\_gaps\_np\_nightsoff0\_delayed-1 &  64  & presto\_gap2\_5 & 106 & vary\_gp\_gpfrac0\_40 \\
23  & long\_gaps\_np\_nightsoff1\_delayed1827  & 65  & presto\_gap3\_0\_mix & 107 & vary\_gp\_gpfrac0\_45 \\
24  & long\_gaps\_np\_nightsoff1\_delayed-1 & 66 & presto\_gap3\_0 & 108 & vary\_gp\_gpfrac0\_50 \\
25  & long\_gaps\_np\_nightsoff2\_delayed1827  & 67  & presto\_gap3\_5\_mix & 109 & vary\_gp\_gpfrac0\_55 \\
26  & long\_gaps\_np\_nightsoff2\_delayed-1 & 68  & presto\_gap3\_5 & 110 & vary\_gp\_gpfrac0\_75 \\
27  & long\_gaps\_np\_nightsoff3\_delayed1827 & 69  & presto\_gap4\_0\_mix & 111 & vary\_gp\_gpfrac1\_00 \\
28  & long\_gaps\_np\_nightsoff3\_delayed-1 & 70 & presto\_gap4\_0 & 112 & vary\_nes\_nesfrac0\_01  \\
29  & long\_gaps\_np\_nightsoff4\_delayed1827 & 71  & presto\_half\_gap1\_5\_mix &         113 & vary\_nes\_nesfrac0\_05  \\
30  & long\_gaps\_np\_nightsoff4\_delayed-1 & 72 & presto\_half\_gap1\_5 & 114 & vary\_nes\_nesfrac0\_10 \\
31  & long\_gaps\_np\_nightsoff5\_delayed1827 & 73  & presto\_half\_gap2\_0\_mix & 115 & vary\_nes\_nesfrac0\_15  \\
32  & long\_gaps\_np\_nightsoff5\_delayed-1 &  74  & presto\_half\_gap2\_0  & 116 & vary\_nes\_nesfrac0\_20   \\
33  & long\_gaps\_np\_nightsoff6\_delayed1827  & 75 & presto\_half\_gap2\_5\_mix & 117 & vary\_nes\_nesfrac0\_25  \\
34  & long\_gaps\_np\_nightsoff6\_delayed-1 &  76  & presto\_half\_gap2\_5  & 118 & vary\_nes\_nesfrac0\_30  \\
35  & long\_gaps\_np\_nightsoff7\_delayed1827  & 77  & presto\_half\_gap3\_0\_mix &        119 & vary\_nes\_nesfrac0\_35   \\
36  & long\_gaps\_np\_nightsoff7\_delayed-1 &  78  & presto\_half\_gap3\_0 & 120 & vary\_nes\_nesfrac0\_40   \\
37  & long\_u1 &  79  & presto\_half\_gap3\_5\_mix & 121 & vary\_nes\_nesfrac0\_45 \\
38  & long\_u2 & 80  & presto\_half\_gap3\_5 & 122 & vary\_nes\_nesfrac0\_50 \\
39  & roman  &  81  & presto\_half\_gap4\_0\_mix & 123 & vary\_nes\_nesfrac0\_55  \\
40  & virgo\_cluster & 82  & presto\_half\_gap4\_0 &  124 & vary\_nes\_nesfrac0\_75 \\
41  & too\_rate50 & 83  & baseline\_retrofoot & 125 & vary\_nes\_nesfrac1\_00   \\
\hline
\end{longtable*}

\begin{figure*}
    \centering
    \includegraphics[scale=0.2]{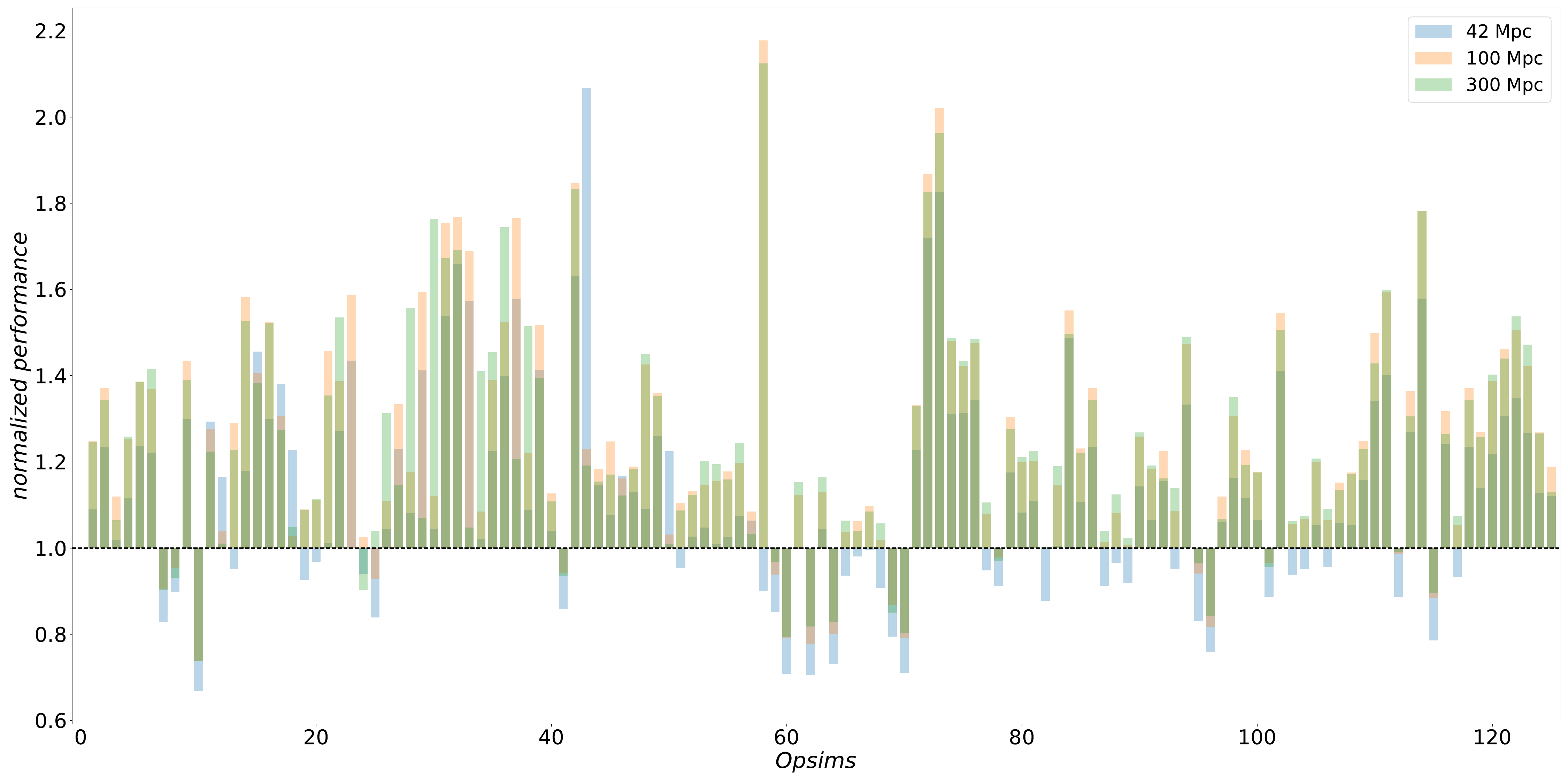}\quad\quad
    \includegraphics[scale=0.2]{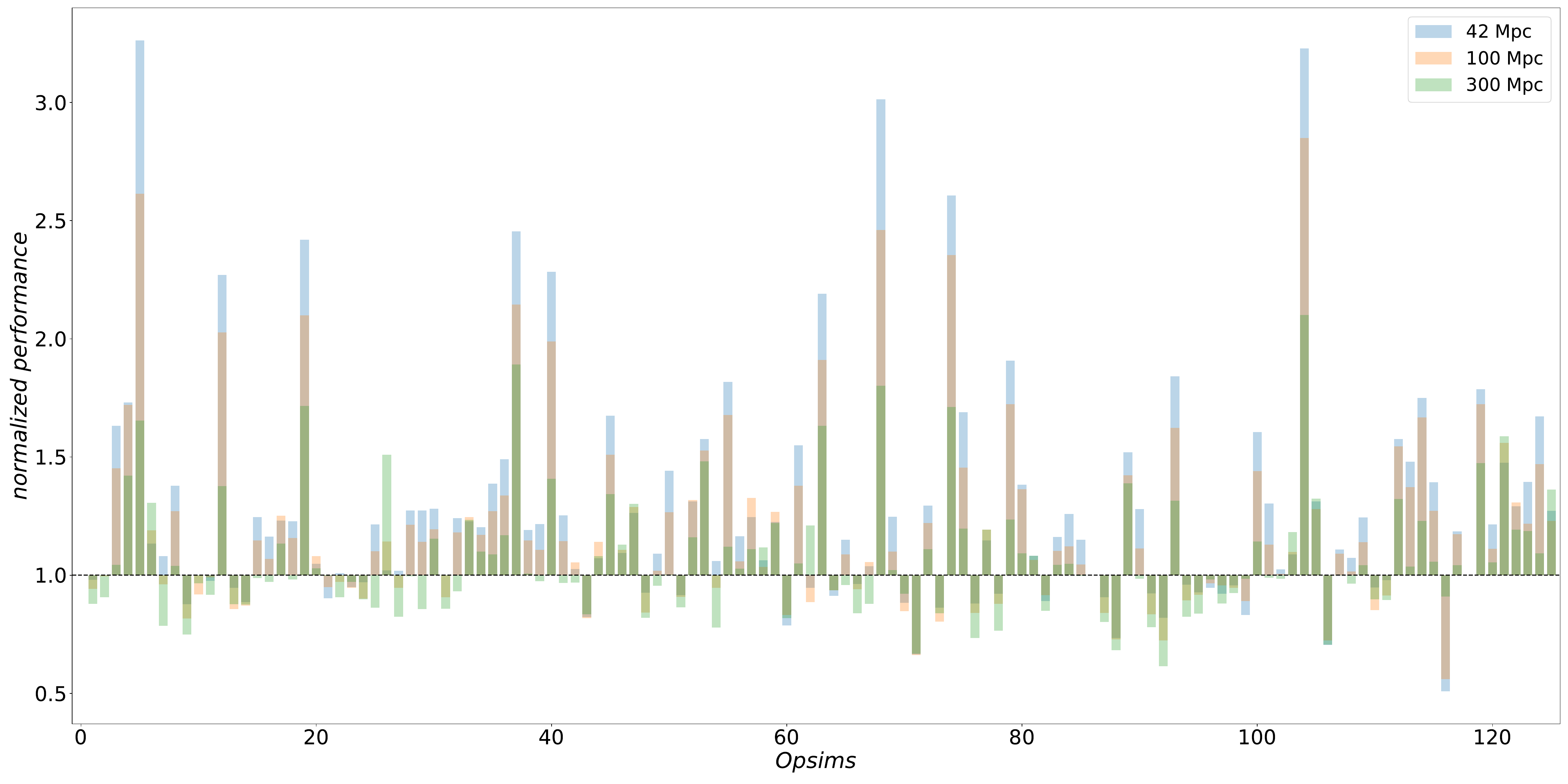}\quad\quad
    \includegraphics[scale=0.2]{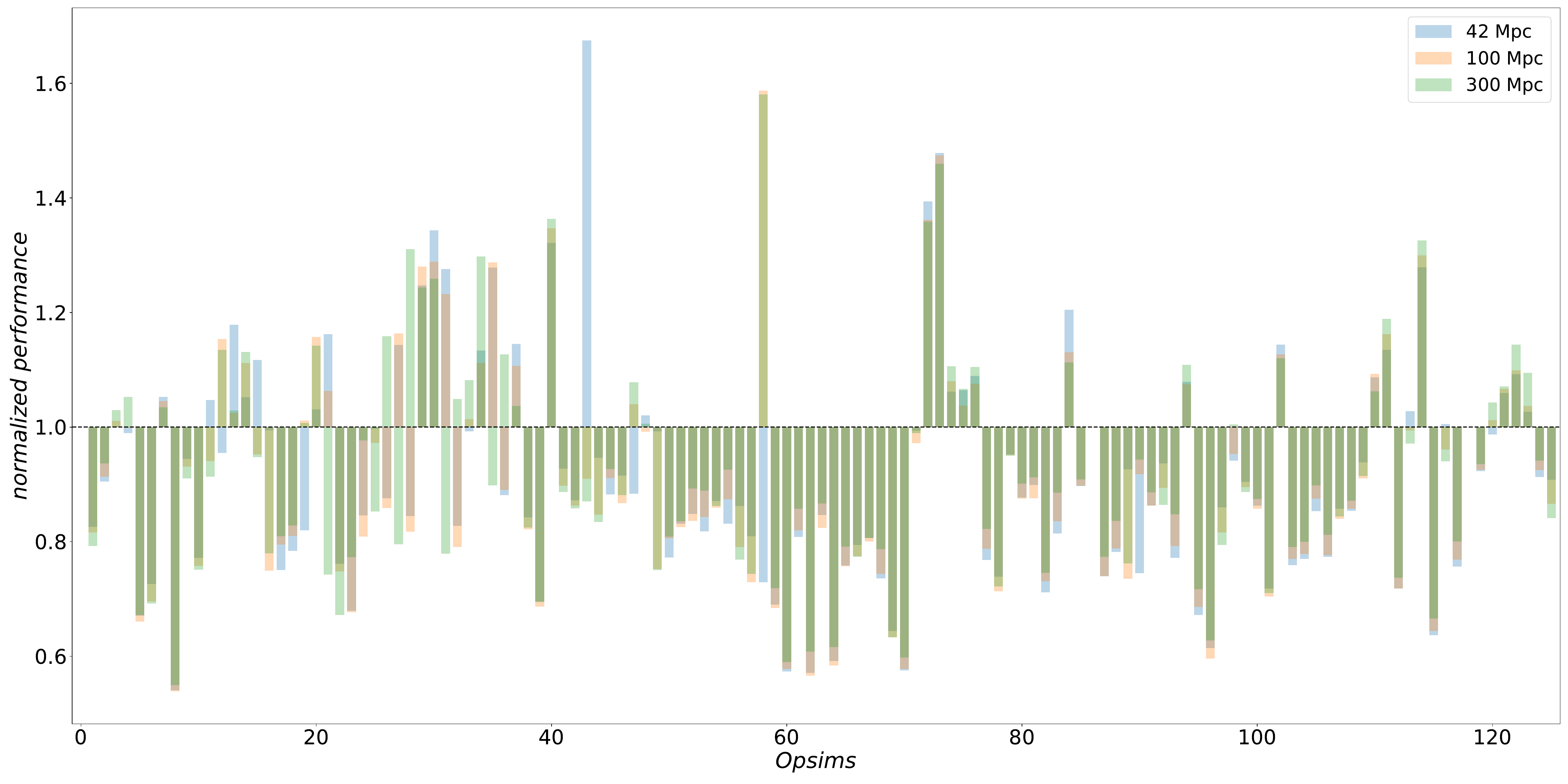}
    \caption{Comparison plot of all the v2.0 \opsim s for a targeted search in a fixed pointing in the Sky, the performance is normalized with respect to the baseline, $N_{balseline}= 7$.
    Each plot consider a population of KN+Afterglow simulated with a fixed viewing angle, from top to bottom the viewing angle is $\theta_{v} = 0, \frac{\pi}{4}, \frac{\pi}{2}$.
    The metric count the median number of detections per filter, which is used as a proxy to evaluate the strategy that will allow the most accurate parameters estimation, as described in \autoref{sec:performance}. Description of \opsim ~indexes shown in the x-axis are reported in \autoref{table:opsim_names}.}
    \label{fig:opsimcomparison_single}
\end{figure*}

\begin{figure*}
    \centering
    \includegraphics[scale=0.25]{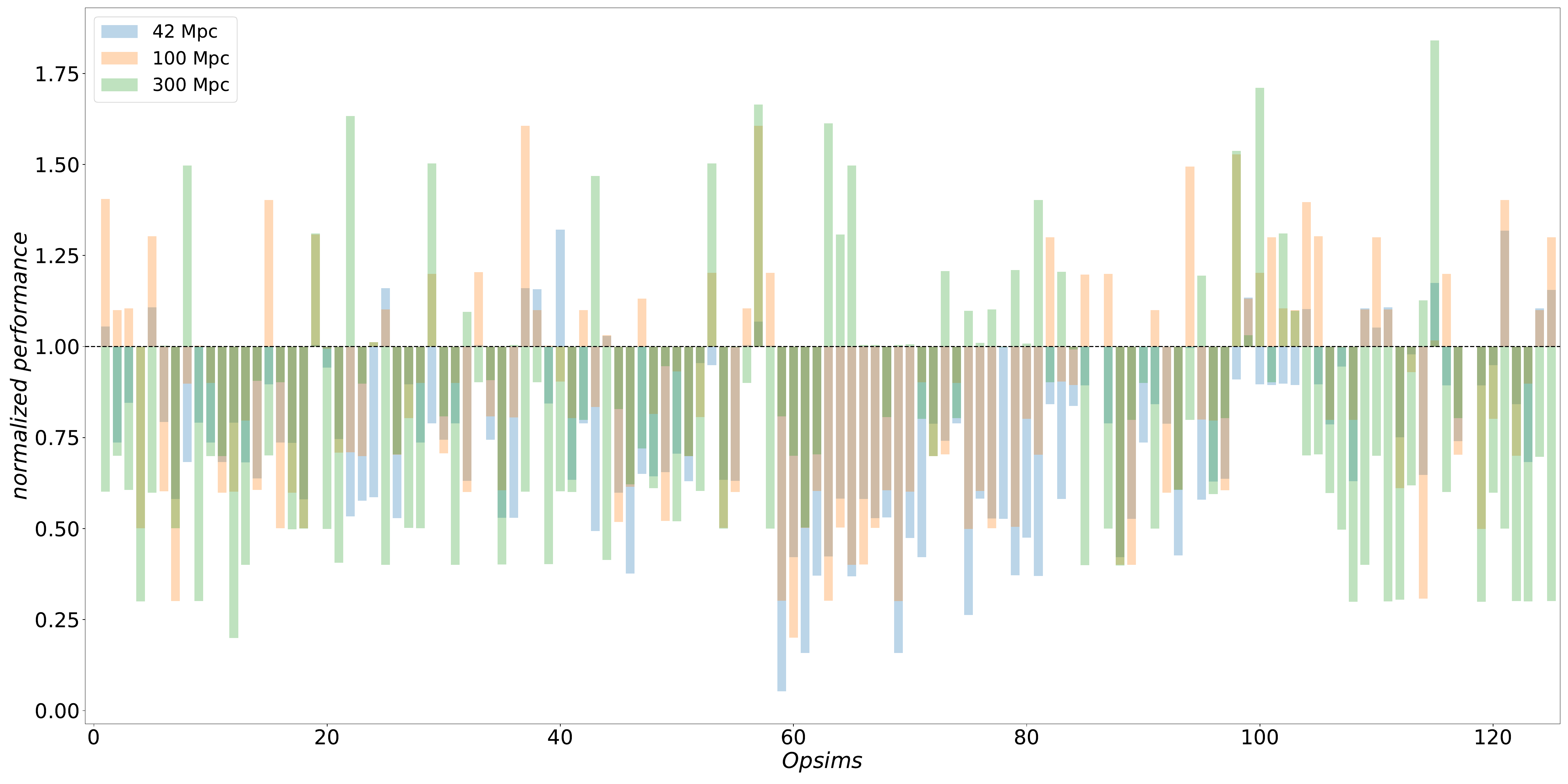}
    \caption{Comparison plot of all the v2.0 \opsim s for an all Sky search, the performance is normalized with respect to the baseline. The metric count the median number of detections per filter, which is used as a proxy to evaluate the strategy that will allow the most accurate parameters estimation, as described in \autoref{sec:performance}. Description of \opsim ~indexes shown in the x-axis are reported in \autoref{table:opsim_names}.}
    \label{fig:opsimcomparison}
\end{figure*}

To support this we compare, among the \opsim s, the mean number of detection per filter (\ie that the higher this number the better the accuracy, see \autoref{fig:opsimcomparison}).

The results show that the \opsim s that allow re-visits within the same nights have higher fraction of well populated light curves, which imply an higher ability to constrain the uncertainties of the model parameters. 
The take away message from this work is that from surveys observations we can expect to improve our detection ability, because we observe deeper and wider, changing from time to time the filters configuration. However, to efficiently constrain the model's parameters we need to maximize the information content of our observations, improving the number of filters and the number of detections we have to observe the evolution of the event.

\section{Discussion and Conclusions}\label{sec:conclusion}
\new{This work aims to understand whether the LSST observing strategy can help collect data that will improve our understanding of KN sources. In \citet{2021Carracedo} they analyzed how to optimize the strategy for filters distribution and survey depth to boost the detection efficiency for these faint and fast-evolving transients. They explore the dependence on the mass of the ejecta, the geometry, the viewing angle, the wavelength coverage, and the source distance. Eventually they claim that the detection efficiency has a strong viewing-angle dependence, especially for filters blueward of i-band. This loss of sensitivity can be mitigated by early, deep observations. Efficient searches for the $gri$ counterpart for KNe\ at $\sim 200$ Mpc would require reaching a limiting magnitude $mag_{lim}=23$ mag within 5 days from explosion, to ensure good sensitivity over a wide range of the model phase-space. Toward this end, \citet{2022AndreoniB} analyze different choices of filter setting and exposure time, and they find that observations in redder $izy$ bands are crucial for identification of nearby (within 300 Mpc) KNe\ that could be spectroscopically classified more easily than more distant sources. LSST’s potential for serendipitous KN discovery could be improved by increasing the efficiency with the use of individual 30s exposures (as opposed to $2\times15s$ snap pairs), with the addition of red-band observations coupled with same-night observations in the g or r bands, and possibly with the further development of a new rolling-cadence strategy.
However, even if detected, the KN are not enough sampled to allow parameters' estimation without ancillary data. }

This work showed how to constrain the parameters of the KNe model using data from the LSST-Vera Rubin Observatory. This new facility is expected to push forward the knowledge of the physics of compact object, improving the statistics of unique transient events such as KNe \citep{Andreoni_2020}. However, to be able to deeply comprehend the compact objects EoS and thermalization processes \citep{Korobkin:2012uy, 2016Barnes}, together with the energy-dependent photon opacities in r-process matter \citep{2020Tanaka, Even_2020}, constraining the uncertainties originated form various assumptions in the modelling is essential. This is due to the complexity of the underlying physics, which is affected by a diverse interactions and scales \citep[see][and reference therein]{2017Metzger}.

The possibility to extract information about the source of a KN events depends on a number of assumptions, among those:
\begin{itemize}
    \item the KN model;
        \vspace{-0.3cm}
    \item the available filters for the observation;
        \vspace{-0.3cm}
    \item the distance;
        \vspace{-0.3cm}
    \item the time window the event is above the observation limit.
\end{itemize}

The search of the KNe light curve can be obtained through discovery of the transient during searches on the entire probability distribution map from the GW trigger or as a targeted search in a small number of specific and very limited regions of the Sky. In the following section we refer to those two scenario as "All Sky search" and "Targeted search". 

We analyzed the possibility to constrain the ejecta mass, velocity and opacity from the photometric multi-wavelength search of KN events assuming we know its distance and position from other messengers (i.e. GW, GRB). The parameter estimation of a transient lightcurve, however, is something that is usually done after the follow up, so the ansatzes here are that we already took care about distance estimation, contaminants and candidate selections, thus we are only interested in analyzing how the search design impacts the parameters estimation.

There is a trade-off between the fitting performance and the features of the observed light curves. In order to be able to extract precise fits of a model's parameters some conditions of measurability need to be satisfied. They can be summarized as follows: 

\begin{itemize}
    \item recognize the different phases of the light curve evolution;
        \vspace{-0.3cm}
    \item gather observation in at least 3 filters.
\end{itemize}

The criteria listed above can be translated to survey strategy design: higher re-visits (with the best strategy considering long gap pairs of about 2 hours every 2 nights) in the same or different filters will weight differently in the observing strategy in the two scenario (All-Sky and targeted search). We use as proxy of the performance the average number of detection per filter (see \autoref{allsky} for details on the Figure of Merit, FoM). We, then, normalize the FoM with respect the FoM for the \texttt{baseline}, $N_{balseline}= 7$ for both "All Sky search" and "Targeted search".
 
For targeted searches, the possibility of considering at least one of the two criteria perform better with respect to the \texttt{baseline}, as visible in  \autoref{fig:opsimcomparison_single}. Eventually the configuration of ToO along the main strategy will improve our ability to extract precise information using only photometric light curve to constrain the source source's ejecta parameters. 

{As shown in middle panel of \autoref{fig:opsimcomparison_single}, when a targeted search is considered \texttt{presto\_gap} (IDs =  63, 68, 74) and \texttt{long\_gaps\_np} (IDs = 5, 12, 19, 37, 40) \opsim s almost double the performance of the \texttt{baseline} in constraining the model parameters for farther sources, this is due to the color information these strategies allow to obtain. Indeed, the \texttt{presto\_gap\_half} add a third visit within the same night for half the nights of the survey, with variations on the time interval between the first pair of visits (standard separation of 33 minutes) to the third visit. Among this family the best strategy is \texttt{presto gap3.5} which consider triples spaced 3.5 hours apart ($g+r$,$r+i$,$i+z$ are the initial pairs). 
\texttt{long\_gaps\_np}, similarly, extends the gap between the pair of visits, modifying it to a variable time period of between 2 to 7 hours. The pair of visits are both in the same filter, in any of griz ($g+r$, $r+i$, or $i+z$ pairs). In some of the simulations, these long gap visits are obtained throughout the survey, while for other simulations the longer time separations do not start until year 5. Among this family of \opsim s the best performing strategy is \texttt{long\_gaps\_nightsoff0}, which considers long gap pairs every night. When we analyze farther sources it appears that having long gap pairs every 4 to 7 nights allow to constrains better the model parameters being \texttt{long\_gaps\_np\_nightsoff7} the best performing \opsim for this family in case of source at 300 Mpc. Overall the best performing \opsim~ is \texttt{vary\_gp\_gpfrac0.30} (ID = 104), \texttt{vary\_gp} family is a set of simulations that investigate the effect of varying the amount of survey time spent on covering the background (non-WFD-level) Galactic Plane area. The combination of image quality, set of filters per observation and cadence allow the best coverage for our simulated light curves.}

\new{Across the different panels in \autoref{fig:opsimcomparison_single} the difference in performances when the population of event is considered at different viewing angles (from top to bottom,$\theta_{v} = 0,\frac{\pi}{4}, \frac{\pi}{2}$) can be analysed. The general discussion still hold, however we see that a worsening of the performance is evident. 
Whereas \texttt{long\_gaps\_np} family appear to outshine the baseline in all the cases. Pointing to the importance to revisit in the same night with different filters pairs to have a well constrained characterization of the events.}

When an All-Sky search (i.e. a search on the entire probability distribution map from the GW trigger) is considered the main criterion for a better performance appears to be the homogeneity of the filter coverage (see \autoref{allsky}), meaning that strategies which respect the criteria in a higher number of region perform better then the \texttt{baseline}. From the results (see \autoref{fig:opsimcomparison}) it is shown that the \texttt{baseline} performs better then almost all the \opsim s, the ones with higher performance differs from this strategy in varying the exposure time or the number of images per exposure to force the limiting magnitude to be homogeneous all over the Sky \citep[see the \texttt{vary} family details,][]{2009LSST}\footnote{See also \url{https://pstn-053.lsst.io/}}.

Indeed, the best performing \opsim~is \texttt{multi\_short}, which takes 4 short (5s) visits per filter in a row, and it stops after 12 short visits per filter in a year and it achieves $\sim$ 700 visits per pointing.

In short, the \texttt{baseline} is a great compromise among all the strategies for the KNe science, and in future an improvement of the ability to constrain parameters of serendipitous discovery of KN events is also foreseen. However, small changes of this strategy oriented to add third image for color information within the 4 hour gap or ad-hoc ToO strategies to follow up the light curve evolution will enhance by a factor of 2 the targeted search ability to describe the KNe events with the most reliable KN models known up to now. 

\section{Acknowlegement}
This work was initially thought out and developed at The Unconventional Thinking Tank Conference 2022, which is supported by INAF. FR thanks Om Shalam Salafia for his very useful comments that improved the manuscript.
This work was supported by the Preparing for Astrophysics with LSST Program, funded by the Heising Simons Foundation through grant 2021-2975, and administered by Las Cumbres Observatory.
MWC acknowledges support from the National Science Foundation with grant numbers PHY-2010970 and OAC-2117997.
Fabio Ragosta thanks the LSSTC Data Science Fellowship Program, which is funded by LSSTC, NSF Cybertraining Grant \#1829740, the Brinson Foundation, and the Moore Foundation; his participation in the program has benefited this work.

\newpage
\appendix
\section{Detection Rate of LSST}\label{TableRate}
In order to computed the detection rates of LSST reported in Sec.\ref{sec:det_rate}, we followed the methodology described in \cite{2022Colombo, 2023Colombo}, starting from a population of merging binary neutron stars (BNS) with a power-law chirp mass and mass ratio probability distribution, fitted to the constraints from both Galactic BNSs and the GW-detected binaries GW170817 and GW190425. The cosmic merger rate density was computed by convolving a $t_d^{-1}$ delay time distribution with a minimum delay time of $t_{d,min}=50$ Myr and the cosmic star formation rate from \citet{2014Madau}, normalized to a local rate density of $R_0=\ensuremath{347_{-256}^{+536}\,\mathrm{Gpc^{-3}\,yr^{-1}}}$ \citep{2022Colombo}. 

For each event we evaluated the GW signal-to-noise-ratio (S/N) through the \texttt{GWFAST} software package \citep{2022Iacovelli}, using the \texttt{IMRPhenomD$\_$NRTidalv2} waveform model \citep{2019Dietrich}. We assumed a network consisting of the two aLIGO, Advanced Virgo and KAGRA, considering the projected O4 sensitivities (the highest targets) and a $70\%$ uncorrelated duty cycle for each detector.

For every BNS, assuming the SFHo equation of state \citep{2013Steiner}, we computed the expected ejecta mass, ejecta average velocity and accretion disk mass using fitting formulae from \cite{2020Barbieri,2020Kruger,2018Radice}. Using this information we evaluated the KN lightcurves from 0.1 to 50 days in the $y$, $z$, $i$, $r$, $g$ and $u$ bands using the model from \cite{2017Perego,2021Breschi}, see Appendix B2 in \cite{2022Colombo} for further explanation. 

For mergers whose remnants promptly collapse or transition to a black hole after a short-lived hypermassive NS phase ($M_\mathrm{rem}\geq 1.2 M_\mathrm{TOV}$), it was assumed that the system launches a relativistic jet, with the energy determined by the mass of the accretion disk and the spin of the remnant. In cases where the jet energy exceeds a threshold defined in \citet{2018Duffell}, the relativistic jet is expected to break out of the ejecta cloud and produce GRB prompt and afterglow emission, indicating a successful jet. For these systems we assumed the jet angular structure of GRB170817A \citep{2019Ghirlanda} and we evaluated the the afterglow lightcurves from 0.1 to 1000 days in the optical ($g$) band. For more information see Appendix B3 from \cite{2022Colombo}.

\section{Bayesian sampler}\label{sampler}

According to Bayes' theorem we compute the posterior probability distribution $p(\bold{\theta}|d,M)$ for the model source parameter $\theta$ and the hypothesis M with data d as:
\begin{equation}
    p(\bold{\theta}|d,M) = \frac{p(d|\bold{\theta},M)p(\bold{\theta}|M)}{p(d|M)} \rightarrow \mathcal{P}(\bold{\theta})= \frac{\mathcal{P}(\bold{\theta})\pi(\bold{\theta})}{\mathcal{Z}(\bold{\theta})}
\end{equation}
where $\mathcal{P}(\bold{\theta}),\pi(\bold{\theta}),\mathcal{Z}(\bold{\theta}) $are the posterior, likelihood, prior, and evidence, respectively. The evidence is the integral taken over the entire domain $\Omega_{\theta}$ of $\bold{\theta}$(i.e. over all possible ):
\begin{equation}
    \mathcal{Z} = \int_{\Omega_{\theta}} \mathcal{L}(\bold{\theta})\pi(\bold{\theta}) d\bold{\theta}
\end{equation}.

\citet{Skilling2004} first conceived and developed nested sampling as a method for estimating Bayesian evidence. The fundamental idea is to integrate the prior in layered "shells" with constant likelihood in order to mimic the evidence. Nested Sampling simultaneously estimates the evidence and the posterior, in contrast to Markov Chain Monte Carlo (MCMC) techniques that can only produce samples proportionate to the posterior. 
Nested Sampling has two main main theoretical requirements:
\begin{enumerate}
    \item Samples must be evaluated sequentially subject to the likelihood constraint $\mathcal{L}_{i+1}>\mathcal{L}_{i}$.
    \item All samples used to compute/replace points in the parameters space must be independent and identically distributed (i.i.d.) random variables drawn from the prior.
\end{enumerate}

The first requirement is entirely algorithmic and straightforward to satisfy (even when sampling in parallel). The second requirement, however, is much more challenging if we hope to sample efficiently: while it is straightforward to generate samples from the prior, by design Nested Sampling makes this simple scheme increasingly more inefficient since the remaining prior volume shrinks exponentially over time.

Solutions to this problem often involve some combination of:
\begin{enumerate}
    \item Proposing new positions by 'evolving' a copy of one (or more) current position to new (independent) positions subject to the likelihood constraint, and
    \item Bounding the iso-likelihood contours using simple but flexible functions in order to exclude regions with lower likelihoods.
\end{enumerate}

In general, the contribution to the posterior at a given value (position) $\bold{\theta}$ has two components. The first arises from the particular value of the posterior itself,$P(\bold{\theta}$). The second arises from the total (differential) volume $dV(\bold{\theta})$
 encompassed by all $\bold{\theta}$’s with the particular $P(\bold{\theta})$. We can understand this intuitively: contributions from a small region with large posterior values can be overwhelmed by contributions from much larger regions with small posterior values.

The compromise between the two elements means that the regions which contribute the most to the overall posterior are those that maximize the joint quantity
\begin{equation}
    w(\bold{\theta})\approx P(\bold{\theta}dV(\bold{\theta})
\end{equation}

This region typically forms a 'shell' surrounding the mode (i.e. the maximum a posteriori (MAP) value) and is what is usually called the typical set. This behavior becomes more accentuated as the dimensionality increases: since volume scales as $r^D$, increasing the dimensionality of the problem creates exponentially more volume farther away from the posterior mode.

Unlike MCMC or similar methods, Nested Sampling starts by randomly sampling from the entire parameter space specified by the prior. In addition to affecting the evidence estimate, the prior also directly affects the overall expected runtime. Indeed, because the parameter space volume increases as $r^D$, increasing the size of the prior directly impacts the amount of time needed to integrate over the posterior; thus, priors should be carefully chosen to ensure models can be properly compared using the evidences computed from Nested Sampling. 

\section{Figure of merit of KN parameters estimation}\label{allsky}
Parameters estimation is a task that is influenced by many factor, some due to the data: \begin{itemize}
    \item number of points on the light curve,
    \item number of filters available,
    \item coverage of the light curve evolution,
\end{itemize}

others due to the algorithm:
\begin{itemize}
    \item cost function,
    \item choice of priors,
    \item model's uncertainties.
\end{itemize}

In \autoref{sec:performance} we stated that number of points on the light curve and number of available filters are fundamental hyper-parameters to constrain the performance of the model parameter estimation. 
Because of this results, we summarised the previously listed criteria in a FoM:
\begin{equation}
    FoM = \frac{1}{6}\sum_{f= [u,g,r,i,z,y] } n^f_{points},
\end{equation}
with,
\begin{equation}
\centering
n^f_{points} = \sum_{t\in \texttt{MJD}}\begin{cases}
1 ~if ~m_f(t)<m_{lim}\\
0 ~if~ m_f(t)>m_{lim}
\end{cases}.
 \end{equation}
where the normalization to $6$ is the total number of filter available for LSST and $m_f(t)$ is the light curve evolution in the ugrizy bands and $m_{lim}$ if the limiting magnitude of the observing night.

\bibliography{refs}

\end{document}